\pdfoutput=1
\documentclass[PAPER, atlasdraft=false, cernpreprint,texlive=2023, USenglish, abbrevref=false, texmf, orcidlogo]{atlasdoc}
\usepackage{atlaspackage}
\usepackage{atlasbiblatex}

\usepackage[jetetmiss, BSM=true, misc=true]{atlasphysics}
\graphicspath{{logos/}{figures/}}

\usepackage{ANA-SUSY-2018-25-PAPER-defs}
\newenvironment{DIFnomarkup}{}{}

\AtlasTitle{Search for supersymmetry in final states with missing transverse momentum and
charm-tagged jets using 139 fb$^{-1}$ of proton--proton collisions at $\sqrt{s}=13$~TeV with the ATLAS detector}

\AtlasJournalRef{JHEP 02 (2025) 193}
\AtlasDOI{10.1007/JHEP02(2025)193}
\AtlasAbstract{%
The paper presents a search for supersymmetric particles produced in
proton--proton collisions at $\sqrt{s}=13$~TeV and decaying into final states with missing
transverse momentum and jets originating from charm quarks. The data were taken with the
ATLAS detector at the Large Hadron Collider at CERN from 2015 to 2018
and correspond to an integrated luminosity of 139 fb$^{-1}$. No
significant excess of events over the expected Standard Model
background expectation is observed in optimized signal regions, and limits are set on the production cross-sections of the supersymmetric particles.
Pair production of charm squarks or top squarks, each decaying into a
charm quark and the lightest supersymmetric particle $\tilde{\chi}^0_1$, is excluded
at 95\% confidence level for squarks with masses up to 900~GeV for scenarios where the mass of $\tilde{\chi}^0_1$ is below 50~GeV. Additionally, the production of leptoquarks with masses up to 900~GeV is excluded for the scenario where up-type leptoquarks decay into a charm quark and a neutrino. Model-independent
limits on cross-sections and event yields for processes beyond the Standard Model are also reported.
}

\AtlasRefCode{SUSY-2018-25}

\PreprintIdNumber{CERN-EP-2024-218}

\AtlasJournal{JHEP}
\hypersetup{pdftitle={ATLAS document},pdfauthor={The ATLAS Collaboration}}
\begin{document}

\maketitle

\tableofcontents
\clearpage
%
%

% The next lines are included from the .//tex/introduction.tex input file
\section{Introduction}

\par Supersymmetric (SUSY) extensions of the Standard Model (SM) of particle physics predict new particles which are partners of SM
particles~\cite{Golfand:1971iw,Volkov:1973ix,Wess:1974tw,Wess:1974jb,Ferrara:1974pu,Salam:1974ig} -- the partners of SM fermions being bosons and those of SM bosons being fermions -- and provide solutions to the hierarchy problem~\cite{Sakai:1981gr,Dimopoulos:1981yj,Ibanez:1981yh,Dimopoulos:1981zb}. $R$-parity-conserving SUSY models~\cite{Farrar:1978xj} are considered in this paper and predict that the supersymmetric partners of quarks (squarks) will be produced in pairs at the Large Hadron Collider (LHC) and that the lightest SUSY particle (LSP) is stable. The lightest neutralino, \ninoone, is considered to be the LSP and is a dark-matter candidate~\cite{Goldberg:1983nd,Ellis:1983ew}.

\par
\begin{figure}[!b]
\centering
\subfloat[]{\label{fig:signal_feynman_a}\includegraphics[width=0.3\textwidth]{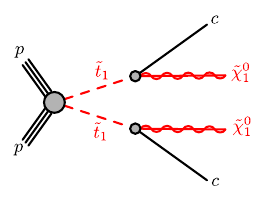}}
\subfloat[]{\label{fig:signal_feynman_b}\includegraphics[width=0.3\textwidth]{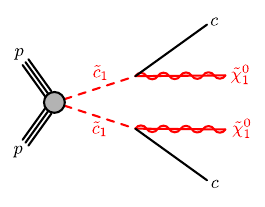}}
\subfloat[]{\label{fig:signal_feynman_c}\includegraphics[width=0.38\textwidth]{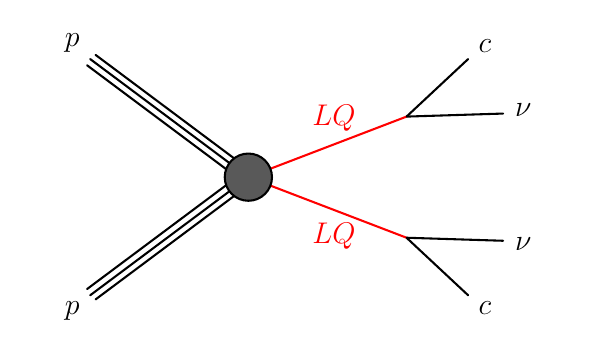}}
\caption{Representative diagrams for pair production of (a) top squarks (\stopone) and (b) charm squarks (\scharmone), with subsequent decay into
charm quarks and neutralinos (\ninoone), and (c) pair production of leptoquarks decaying into charm quarks and neutrinos.}
\end{figure}

 The search described in this paper targets SUSY models where top or charm squarks (\stopone/\scharmone) are the lightest squarks, with masses less than about one~\TeV~\cite{Mahbubani:2012qq}.
For top squarks considered in this search, flavor-violating effects allow the top squark to decay into a charm quark and an LSP, as shown in
\cref{fig:signal_feynman_a}. This results in a final state with two charm quarks and missing transverse momentum from the LSPs, which escape detection. The extent
of flavor-violating effects is model dependent, but when there is significant mixing between the top squark and the charm squark~\cite{Grober:2015fia} which can be due to renormalisation group equations, even if minimal flavor violation is
assumed at a certain scale, the branching
fraction of $\stopone\to c \LSP$ can exceed the branching fraction of $\stop\to t^*
\LSP$, where $t^*$ can be either an on-shell or off-shell top quark.

\par In supersymmetric models with minimal flavor-violating effects, charm squarks could be considerably lighter than other squarks. This motivates searches for pair production of charm squarks, with each charm squark decaying into a charm quark and an LSP as shown in \cref{fig:signal_feynman_b}. Notably, the model of charm squark pair production in \cref{fig:signal_feynman_b} shares the same experimental signature and production cross-section as for the top squark pairs in \cref{fig:signal_feynman_a}. The
experimental signature includes jets that contain charm hadrons but no bottom hadrons and momentum imbalance
caused by the two LSPs. The search discussed in this paper was optimized for the case
where the branching fraction of the decay of top or charm squark into a charm quark plus the
LSP is unity. However, the results of the search are also interpreted in the scenario of top
squark production with a varying branching fraction of the top squark decay into a charm quark
plus the LSP ranging from 0.1 to 1, with the alternative decay being to a top quark plus the LSP.

\par Leptoquark (LQ) models~\cite{DIMOPOULOS1979237,DIMOPOULOS198069,EICHTEN1980125,ANGELOPOULOS198759,BUCHMULLER1986377,PhysRevD.10.275,PhysRevLett.32.438,Diaz_2017,Belyaev_2005}
with up-type scalar LQs (\lqu) or vector LQs (vLQ) predict their pair production and subsequent decay into $c\nu c\nu$, with the same
experimental signature as in top and charm squark production models (see \cref{fig:signal_feynman_c}). The following models are considered: $U_1$
vLQ models that couple to second-generation quarks and third-generation leptons (\vlq, where the subscript indicates the quark and lepton generations)
in both the minimal and Yang--Mills coupling
scenarios~\cite{Baker:2019sli,Blumlein:1996qp}, assuming a branching fraction
($\mathcal{B}$) of $\mathcal{B}(\vlq\to c\nu_\tau)=0.5$ in both cases with the
alternative decay being $\vlq\to s\tau$; and
models where scalar LQs couple to second-generation quarks and first- or second-generation leptons~\cite{Mandal:2015lca,Kramer:2004df,Kramer:1997hh}, allowing a range of
$\mathcal{B}(\lqu\to c \nu_{e/\mu})$ values (here the alternative decay
is $\lqu\to s e/\mu$).

\par This search for top squark, charm squark, and leptoquark pair production uses $pp$ collisions from the LHC at a centre-of-mass energy of $\sqs=\qty{13}{\TeV}$ recorded by the ATLAS experiment from 2015 to 2018. A previous search by ATLAS~\cite{SUSY-2016-26}, based on the identification of jets containing charm hadrons but no bottom hadrons ($c$-tagging), used $\qty{36.1}{\ifb}$ of collisions and was able to probe top and charm squark masses up to \qty{850}{\GeV}. In addition to a larger dataset, the search herein also benefits from improvements in $c$-tagging and from an advanced technique, Recursive Jigsaw Reconstruction~\cite{Jackson:2017gcy} which makes use of jets originating from initial-state radiation (ISR), and which provides sensitivity to models with small mass splittings between the top squark and the LSP. Other ATLAS results also have sensitivity to the models considered in this paper: a search using final states with an energetic jet (but no $c$-tagging) and large missing transverse momentum~\cite{EXOT-2018-06} was particularly sensitive to models with small mass differences between the top/charm squark and the LSP, while a search for single squark production that did not utilize $c$-tagging~\cite{SUSY-2018-22} had good sensitivity to models with large top/charm squark masses, and a more recent search for top squark pair production focused on squarks decaying into either a top quark or charm quark plus an LSP~\cite{tcMET}. The most recent search performed by the CMS
Collaboration~\cite{CMS-SUS-19-010} used $\qty{137}{\ifb}$ of $\sqs=\qty{13}{\TeV}$ $pp$ collisions from the LHC, but only considered mass differences between the
top squark and the LSP of less than 80~GeV, while the present ATLAS search explores mass differences ranging from 20~GeV to about 1~TeV. Finally, for the considered leptoquark models, this paper is the first to report results for the described final state from an LHC experiment.

% End of text imported from the .//tex/introduction.tex input file

% The next lines are included from the .//tex/atlas-detector.tex input file
\section{ATLAS detector}

\newcommand{\AtlasCoordFootnote}{ATLAS uses a right-handed coordinate system with its origin at the nominal interaction point (IP)  in the centre of the detector and the \(z\)-axis along the beam pipe.  The \(x\)-axis points from the IP to the centre of the LHC ring,  and the \(y\)-axis points upwards.  Polar coordinates \((r,\phi)\) are used in the transverse plane,  \(\phi\) being the azimuthal angle around the \(z\)-axis.  The pseudorapidity is defined in terms of the polar angle \(\theta\) as \(\eta = -\ln \tan(\theta/2)\) and is equal to the rapidity  $ y = \frac{1}{2} \ln \left( \frac{E + p_z}{E - p_z} \right) $ in the relativistic limit.  Angular distance is measured in units of \(\Delta R \equiv \sqrt{(\Delta y)^{2} + (\Delta\phi)^{2}}\).}

The ATLAS detector~\cite{PERF-2007-01} at the LHC covers nearly the entire solid angle around the collision point.\footnote{\AtlasCoordFootnote}
It consists of an inner tracking detector surrounded by a thin superconducting solenoid, electromagnetic and hadronic calorimeters,
and a muon spectrometer incorporating three large superconducting air-core toroidal magnets.

The inner-detector system (ID) is immersed in a \qty{2}{\tesla} axial magnetic field
and provides charged-particle tracking in the range \(|\eta| < 2.5\).
The high-granularity silicon pixel detector covers the vertex region and typically provides four measurements per track,
the first hit generally being in the insertable B-layer (IBL) installed before Run~2~\cite{ATLAS-TDR-19,PIX-2018-001}.
It is followed by the SemiConductor Tracker (SCT), which usually provides eight measurements per track.
These silicon detectors are complemented by the transition radiation tracker (TRT),
which enables radially extended track reconstruction up to \(|\eta| = 2.0\).
The TRT also provides electron identification information
based on the fraction of hits (typically 30 in total) above a higher energy-deposit threshold corresponding to transition radiation.

The calorimeter system covers the pseudorapidity range \(|\eta| < 4.9\).
Within the region \(|\eta|< 3.2\), electromagnetic calorimetry is provided by barrel and
endcap high-granularity lead/liquid-argon (LAr) calorimeters,
with an additional thin LAr presampler covering \(|\eta| < 1.8\)
to correct for energy loss in material upstream of the calorimeters.
Hadronic calorimetry is provided by the steel/scintillator-tile calorimeter,
segmented into three barrel structures within \(|\eta| < 1.7\), and two copper/LAr hadronic endcap calorimeters.
The solid angle coverage is completed with forward copper/LAr and tungsten/LAr calorimeter modules
optimised for electromagnetic and hadronic energy measurements respectively.

The muon spectrometer (MS) comprises separate trigger and
high-precision tracking chambers measuring the deflection of muons in a magnetic field generated by the superconducting air-core toroidal magnets.
The field integral of the toroids ranges between \num{2.0} and \qty{6.0}{\tesla\metre}
across most of the detector.
Three layers of precision chambers, each consisting of layers of monitored drift tubes, cover the region \(|\eta| < 2.7\),
complemented by cathode-strip chambers in the forward region, where the background is highest.
The muon trigger system covers the range \(|\eta| < 2.4\) with resistive-plate chambers in the barrel, and thin-gap chambers in the endcap regions.

The luminosity is measured mainly by the LUCID--2~\cite{LUCID2} detector that records Cherenkov light produced in the quartz windows of photomultipliers located close to the beampipe.

Events are selected by the first-level trigger system implemented in custom hardware,
followed by selections made by algorithms implemented in software in the high-level trigger~\cite{TRIG-2016-01}.
The first-level trigger accepts events from the \qty{40}{\MHz} bunch crossings at a rate below \qty{100}{\kHz},
which the high-level trigger further reduces in order to record complete events to disk at about \qty{1}{\kHz}.

A software suite~\cite{SOFT-2022-02} is used in data simulation, in the reconstruction
and analysis of real and simulated data, in detector operations, and in the trigger and data acquisition
systems of the experiment.

% End of text imported from the .//tex/atlas-detector.tex input file

% The next lines are included from the .//tex/data_sim.tex input file
%
\section{Data and simulated event samples}

\par The data used in this analysis were recorded with a fully operational detector from stable-proton-beam collisions, and are clear of a significant amount of cosmic-ray or beam-induced background~\cite{DAPR-2018-01}.
The total integrated luminosity of this dataset is $139.0\pm2.4$~\ifb~\cite{ATLAS-CONF-2019-021}, determined using the LUCID-2
detector~\cite{Avoni:2018iuv} for the primary luminosity measurements, complemented by measurements from the inner detector and calorimeters.

The LHC accelerates and collides protons in bunches, resulting in multiple
$pp$ collisions during each bunch crossing, alongside the much rarer hard-scattering
$pp$ collisions of interest. The bunch crossings happen every 25~ns. The average number of additional proton--proton interactions per bunch crossing (\pileup) grew from 13 in 2015 to a value of about 38 in 2017.

\par Missing transverse momentum triggers were used to select the
data~\cite{TRIG-2019-01} for this analysis. The triggers had
thresholds of 70~GeV in 2015 and early 2016, 100~GeV in late 2016 and
2017, and 110~GeV in 2018. These triggers reach an efficiency plateau
for offline missing transverse momenta greater than \qty{200}{\GeV}.

\par The predictions for signal and background contributions are obtained with Monte Carlo (MC) simulated event samples. To match the LHC beam conditions, these simulated event samples also contain \pileup interactions.
The simulated events were processed through the ATLAS detector
simulation~\cite{SOFT-2010-01} and then subjected to the same
reconstruction algorithms as the data. All samples were processed with
the full \GEANT~\cite{Agostinelli:2002hh} detector model with the
exception of the samples with leptoquarks or direct top or charm
squark production which were processed with
\textsc{Atlfast-II} detector simulations~\cite{ATL-SOFT-PUB-2018-002}. Finally, the simulated samples were
corrected to improve the accuracy of the \pileup modelling, jet momenta, lepton efficiency and momenta, trigger
efficiencies, missing transverse momentum, and $c$-tagging efficiency.

\par Standard Model background samples were produced with various event generators as shown in \cref{tab:mc_samples}. All samples, except the ones that use the \SHERPA\ event generator~\cite{Gleisberg:2008ta}, were processed
with \EVTGEN[1.7.0]~\cite{Lange:2001uf} for the $b$- and $c$-hadron decays. The cross-sections, which were used to normalize the samples, were calculated separately with higher-order corrections in the strong
coupling (\alphas) for all samples except for the diboson production ($pp\rightarrow VV$+jets) sample, where the cross-section calculated by the \SHERPA\ event generator was retained. For the \SHERPA\ sample of single vector bosons produced in association with jets ($V$+jets), virtual electroweak loop-terms were included at next-to-leading-order (NLO) accuracy.

\begin{table}[t]
\footnotesize
\centering
\caption{Overview of the simulated background samples. The simulated
vector bosons, $V$, are massive $W$ and $Z$ bosons. The \NNPDF[3.0nlo]~\cite{Ball:2014uwa} parton distribution functions
(PDF) set is used.}
\begin{tabular}{l|l|l|l|l|l }
\hline
Background process            & Matrix element                     & PDF   & Parton showering        & Underlying & Cross-section  \\
&        generator                               &  accuracy    & and hadronization &event  tune       & calculation accuracy \\
\hline\hline
$pp\rightarrow V$+jets & \SHERPA[2.2.11]~\cite{Gleisberg:2008ta} & NNLO & \SHERPA       & Default & NNLO~\cite{PMGR-2021-01}    \\
$pp\rightarrow \ttbar V$+jets        & \AMCatNLO[2.3.3]~\cite{Alwall:2014hca}        & NLO     & \PYTHIA[8.210]~\cite{Sjostrand:2014zea}  & A14~\cite{ATL-PHYS-PUB-2014-021}     & NLO~\cite{Alwall:2014hca} \\
$pp\rightarrow \ttbar$+jets           & \POWHEGBOX[v2]~\cite{Alioli:2010xd}       & NNLO   & \PYTHIA[8.230]  & A14     & NNLO+NNLL~\cite{Czakon:2013goa,Czakon:2012pz,Czakon:2012zr,Baernreuther:2012ws,Cacciari:2011hy,Czakon:2011xx} \\
$pp\rightarrow t$+jets / $tW$+jets         & \POWHEGBOX[v2]                            & NNLO   & \PYTHIA[8.230]  & A14     & NNLO+NNLL~\cite{Kidonakis:2011wy,Kidonakis:2010ux,Kidonakis:2010tc} \\
$pp\rightarrow VV$+jets            & \SHERPA[2.2.1--2.2.2]~\cite{Bothmann:2019yzt}                   & NNLO & \SHERPA       & Default & NLO  \\
\hline
\end{tabular}
\label{tab:mc_samples}
\end{table}

\par The SUSY event samples were generated with \MGNLO[2.8.1]~\cite{Alwall:2014hca} at
leading order (LO) in the strong coupling constant (\alphas) with up to two additional partons using
the \NNPDF[3.0nlo]~\cite{Ball:2014uwa} parton distribution function
(PDF) set and were showered and hadronized with
\PYTHIA[8.244]~\cite{Sjostrand:2014zea} using the A14 set of tuned parameters (tune) in \PYTHIA~\cite{ATL-PHYS-PUB-2014-021}. Samples with a small mass difference between the top squark and \LSP\ (up to 140~GeV) were generated at a later point in time with \MGNLO[2.9.5] and then interfaced with \PYTHIA[8.306]. The matrix element (ME) calculation was performed at tree level including the emission of up to two additional partons. The matching between the ME and the parton shower was implemented using the CKKW-L prescription~\cite{Lonnblad:2011xx}, with the matching scale set to a quarter of the top/charm squark mass. These samples were also processed with \EVTGEN[1.7.0].

\par The top squark and charm squark signal cross-sections were calculated from pair production of scalar colored particles
and with the gluino assumed to be massive enough to not significantly contribute to the production process. Consequently, each production cross-section depends only on the squark mass.
These cross-sections are computed at approximate next-to-next-to-leading-order (NNLO) accuracy in \alphas
with resummation of next-to-next-to-leading logarithmic (NNLL) soft gluon terms~\cite{Beenakker:2016lwe,Beenakker:1997ut,Beenakker:2010nq,Beenakker:2016gmf}.

\par The samples with pair-produced scalar LQs were generated with matrix elements calculated at NLO accuracy in \alphas
with \MGNLO[2.6.0]~\cite{Alwall:2014hca}, using the method described in Ref.~\cite{Mandal:2015lca}, in which NLO matrix elements~\cite{Kramer:2004df,Kramer:1997hh}
are interfaced to \PYTHIA[8.230]. The total cross-sections
of scalar LQ pair production were computed at approximate NNLO accuracy in \alphas
with resummation of NNLL soft gluon terms~\cite{Beenakker:2016lwe,Beenakker:1997ut,Beenakker:2010nq,Beenakker:2016gmf}. The cross-sections do not include lepton $t$-channel contributions, which are neglected in Ref.~\cite{Mandal:2015lca} and may lead
to corrections at the percent level~\cite{Borschensky:2020hot}.
For vector-LQ pair production, the matrix elements were calculated at LO in \alphas following the model described
in Ref.~\cite{Baker:2019sli}, and events were generated using \MGNLO[2.9.5] in conjunction with \PYTHIA[8.306].
Parton luminosities were provided by the five-flavor scheme
\NNPDF[3.0nlo] PDF set and the underlying event was modelled with the A14 tune in \PYTHIA. These samples were also processed with \EVTGEN[1.7.0].

% End of text imported from the .//tex/data_sim.tex input file

% The next lines are included from the .//tex/objects.tex input file
\section{Event reconstruction}

\par Events are required to have a primary vertex reconstructed from at least two
tracks with transverse momentum $\pt>\qty{0.5}{\GeV}$. If more than one such vertex is found, the one with the
largest sum of squares of transverse momenta of associated tracks is selected as
the hard-scattering collision~\cite{ATL-PHYS-PUB-2015-026}. The other vertices are considered as \pileup.

\par Hadronic jet candidates are reconstructed using the \antikt jet algorithm~\cite{Cacciari:2008gp,Fastjet} with radius
parameter $R=0.4$, using particle-flow objects (PFOs)~\cite{PERF-2015-09} as inputs. PFOs are
charged-particle tracks matched to the hard-scatter vertex with the requirement $|\zzsth| < \qty{2.0}{\milli\metre}$
(where $z_0$ is the longitudinal impact parameter) and calorimeter energy clusters
surviving an energy subtraction algorithm that removes the energy contributions deposited by good quality
tracks from any vertex. Jet energy scale corrections, derived from MC simulation
and data, are used to calibrate the average energies of jet candidates to the scale of their
constituent particles~\cite{JETM-2018-05}. Jets with $\pt>\qty{20}{\GeV}$ and $|\eta| < 4.5$ are used to calculate the missing transverse momentum~\cite{ATLAS-CONF-2018-023}, while jets with $|\eta| < 2.8$ are considered to be \enquote{signal} jets. Jets containing a
large particle momentum contribution from \pileup vertices, as measured by the jet vertex
tagger (JVT) discriminant~\cite{PERF-2014-03}, are rejected if they have $\pt < \qty{60}{\GeV}$, $|\eta| < 2.5$ and
a discriminant value of JVT~$< 0.5$.

\par Jets can be identified as $c$-tagged or $b$-tagged jets if they lie within the inner-detector acceptance of $|\eta| < 2.5$. To identify jets containing $c$-hadrons, the charm-tagging
algorithm DL1r$_c$, optimized for a similar search~\cite{tcMET}, is used. The DL1r$_c$ algorithm is based on the ATLAS DL1r algorithm~\cite{FTAG-2019-07}, which uses a selection of inputs associated with the jet
including information about the impact parameters of ID tracks, the presence of displaced secondary vertices, and the reconstructed flight paths of $b$- and $c$-hadrons inside the jet. The
algorithm provides three probabilities for a jet to either contain $b$-hadrons, $c$-hadrons, or light-flavor hadrons. These three probabilities are combined in a similar way to the
$b$-tagging algorithm, but with fine-tuned parameters specifically optimized for identifying jets containing $c$-hadrons but not $b$-hadrons; the $f_{b}$ parameter was set to 0.28. The DL1r
$b$-tagging algorithm is used to identify $b$-jets which are no longer considered as potential $c$-jet candidates. The DL1r algorithm has 77\% identification efficiency for $b$-jets and
20\% and 0.9\% misidentification probabilities for $c$-jets and light-flavor jets, respectively, evaluated in a sample of simulated SM \ttbar\ events. The $c$-tagging algorithm (referred to
as \enquote{$c$-tagging with $b$-veto}) is tuned to have an efficiency of 20\% by requiring that the final DL1r$_c$ discriminant be greater than 1.315. This corresponds to rejection factors
of 29 for $b$-jets and 57 for light-flavor jets; and a $\tau$-lepton misidentification efficiency of 15\%. These efficiencies and rejections factors are determined with a sample of
simulated $t\bar{t}$ events. Both the simulated $c$-jet identification efficiency and misidentification probabilities are corrected to match those measured in data. The approach employed is
similar to that for $b$-tagging~\cite{FTAG-2020-08,FTAG-2018-01,FTAG-2019-02}. Between \pt\ ranges of 20~GeV and 250~GeV, the scale factors are compatible with unity
for all jet-flavor corrections. For $c$-jets, the systematic uncertainties range from 17\% for $\pt< 65$~GeV to a few percent for higher-\pt\ jets (up to 3~TeV), while for $b$-jets and
light-flavor jets, the uncertainties are 5-7\% and approximately 13\%, respectively, for all \pt\ ranges below 3~TeV. For $b$- and $c$-jets with $\pt>3$~TeV, the uncertainties on the scale
factors are up to 30\%.

\par Electron candidates are constructed from energy deposits in the EM calorimeter matched to an ID track. \enquote{Baseline} electron candidates, which are used to veto events with electrons, must pass a loose likelihood-based identification selection~\cite{EGAM-2018-01} and have $\pt>4.5$~\GeV and $|\eta|<2.47$. The longitudinal impact parameter of the ID track associated with a \enquote{baseline} electron is required to satisfy $|\zzsth|<0.5$~mm. Electrons used for regions rich in the dominant SM background processes with prompt electrons (i.e.\ $W$+jets and $Z$+jets) must meet further selection criteria: $\pt>10$~\GeV, \enquote{Loose} isolation~\cite{EGAM-2018-01}, $|d_0/\sigma(d_0)|<5$ where $d_0$ is the transverse impact parameter and $\sigma(d_0)$ is its uncertainty, and the \enquote{Tight} likelihood-based identification selection~\cite{EGAM-2018-01}. Electrons satisfying these criteria are referred to as \enquote{signal} electrons.

\par Muon candidates are found by combining track segments from the inner detector and muon spectrometer or by extrapolating from the forward
sections of the muon spectrometer, in the region $2.5<|\eta|<2.7$. \enquote{Baseline} muon candidates, which are used to veto events with muons,
are required to pass  \enquote{Medium} identification~\cite{MUON-2018-03} and have $\pt>4$~\GeV, $|\eta|<2.7$, and $|\zzsth|<0.5$~mm.
Muons passing the more stringent \enquote{Loose\_VarRad} isolation~\cite{MUON-2018-03},
$|d_0/\sigma(d_0)|<3$, and $\pt>10$~\GeV requirements are referred to as \enquote{signal} muons.

\par Hadronically decaying $\tau$-leptons are reconstructed from jets with $\pt>10$~\GeV{} and either $|\eta|<1.37$ or $1.52<|\eta|<2.5$~\cite{ATL-PHYS-PUB-2015-045}. In addition, $\tau$-lepton
candidates must have $\pt>20$~\GeV, one or three charged-particle tracks, and tracks with a total electric charge of $\pm$1 times the electron charge. The $\tau$-leptons are also required to meet the \enquote{Loose} criterion of
the neural-network-based identification algorithm~\cite{ATL-PHYS-PUB-2019-033}.

\par To prevent double-counting of electrons as jets, jets within $\Delta R = 0.2$ of an electron are not considered further unless the jet is $b$-tagged; $b$-tagged jets are removed if the
overlapping electron has $\pt>100$~\GeV, otherwise the electron is removed. Muons are not counted if they are within $\Delta R =(0.04+10~\GeV/\pt($jet$))$ of a jet. This criterion is applied to remove muons from decays of $b$- and $c$-hadrons but retain high-\pt\ muons resulting from decays of high-\pt massive particles, which tend to be less isolated than muons from decays of low-\pt\ massive particles. Electrons that share an inner-detector track with a muon are discarded. If a $\tau$-lepton candidate is
found to overlap ($\Delta R < 0.2$) with a muon or electron, the $\tau$-lepton is removed. Jets within $\Delta R = 0.4$ of a $\tau$-lepton are removed. This removal of overlapping objects is only performed for baseline objects.

\par The missing transverse momentum $\vec{p}_{\mathrm{T}}^{\mathrm{miss}}$, with magnitude \met{}~\cite{JETM-2020-03}, is calculated as the negative vector
sum of the transverse momenta of jets, $\tau$-leptons, electrons, and muons. Photons are counted as jets since they are not considered in this search. To account for the underlying event, the missing transverse momentum computation also considers tracks with $\pt>0.5$~\GeV that are associated with
the primary vertex but not with any reconstructed particle or jet.

\par The significance of missing transverse momentum, denoted by \metsig, is the ratio of \met{} to its variance~\cite{ATLAS-CONF-2018-038}.
The variance is calculated using the \pt\ resolution of each component used to evaluate the \met. The correlation between the components is also included.

% End of text imported from the .//tex/objects.tex input file

% The next lines are included from the .//tex/signal_regions.tex input file
\section{Event selection}
The search uses simulated data to design signal regions that have sensitivity to the SUSY models of interest.
Control regions that have high purity in particular background processes are used to estimate major background contributions; their normalizations are determined by the data and are extrapolated to the signal regions using simulation. The control regions are designed to not overlap with the signal regions, to have selection criteria similar to those of the signal regions so as to reduce uncertainties from the extrapolation, and to have only a small contamination from signal processes. In addition to the signal and control regions, validation regions are used to validate the background modelling before examining the recorded data in the signal regions (unblinding). Detailed descriptions of the statistical data analysis and how the SM backgrounds are constrained with a fit are provided in~\cref{sec:result}.

\subsection{Signal regions}

\par To maximize the discovery potential, the search uses two sets of signal regions targeting two drastically different kinematic regions: \textit{High-Mass} and \textit{Compressed}. The signal regions were designed using simulated event
samples to maximize the expected sensitivity. The High-Mass signal
regions are sensitive to leptoquark models and SUSY models with large
leptoquark or top/charm squark masses ($\gtrsim\qty{600}{\GeV}$) and a large difference between the top/charm squark and LSP masses
($\gtrsim\qty{200}{\GeV}$); here \met-based variables are relied upon to separate signal
events from background events. The Compressed signal regions are sensitive to SUSY
models with both smaller top squark masses and compressed spectra,
$\Delta\sigPoint\lesssim\qty{175}{\GeV}$; here the presence of high-momentum jets
originating from ISR and Recursive Jigsaw Reconstruction (RJR)~\cite{Jackson:2017gcy} are used to suppress the SM background. The High-Mass and Compressed
regions do not overlap because the High-Mass regions require the leading jet to be a $c$-tagged jet while the Compressed regions require the leading jet not to be $c$-tagged.

\par The signal models targeted in this search are expected to produce events that have
significant \met\ -- even in the Compressed region for which additional jet activity
boosts the neutralinos resulting in large \met\ -- so all events are
required to have $\met>\qty{250}{\GeV}$. The signal processes do not produce any leptons, so
events containing electrons, muons, or $\tau$-leptons are vetoed. All regions require at least one jet with
$\pt>\qty{250}{\GeV}$. For High-Mass signal regions, all jet-related variables are computed using jets with $\pt>40$~\GeV. At least two jets must be present, and at least one must be identified as a $c$-tagged jet. To reject backgrounds
originating from top-quark-pair ($\ttbar$) decays, events with
$b$-tagged jets are discarded. Finally, the minimum azimuthal angle separation ($\Delta\phi$) between the (up to) four
leading jets and the \metvec\ ($\Delta\phi(\metvec,\vec{p}_{\mathrm{T}} (j_{4}))$) is required to be greater than 0.4 to reject multijet backgrounds, reducing them to negligible levels such that they need not be considered further in this search.

\par Three non-overlapping High-Mass signal regions, named SR-HM1, SR-HM2 and SR-HM3, are defined as shown in \cref{tab:srab_def}. The SUSY scenario with $\mathcal{B}(\stopone/\scharmone\to c+\LSP)=1$ and $\sigPoint=(1000, 1)\,\GeV$ is used to optimize the selection for SR-HM1, while the scenario with $\sigPoint=(700, 400)\,\GeV$ is used to optimize the selection for SR-HM2 and SR-HM3. For Compressed signal regions, all jet-related variables are computed using jets with $\pt>20$~\GeV. All signal models are expected to produce two $c$-tagged jets, so all three regions require the presence of at least two $c$-tagged jets. Additionally, the signal models which are used to optimize the High-Mass signal regions are expected to have highly energetic $c$-tagged jets, motivating the requirement that the leading jet be $c$-tagged and that the subleading jet have a \pt\ ($p_{\mathrm{T}} (j_{2})$) of at least 150~\GeV. Standard Model background contributions with two $c$-tagged jets originating from real charm quarks are expected to be produced by processes such as $g \to c\bar{c}$, where $g$ is a gluon, and \mcc, the invariant mass of the two highest-$\pt$ $c$-tagged jets, is expected to have a falling spectrum. Thus, SR-HM1, SR-HM2, and SR-HM-3 require \mcc\ to be at least 200~\GeV. Different \mcc\ requirements are made in SR-HM2 and SR-HM3 to target signal samples with $\Delta\sigPoint
\sim\qty{300}{\GeV}$ and $\sim\qty{400}{\GeV}$, respectively. An upper bound on \mcc\ was added to SR-HM3 to reject $V$+jets backgrounds where \mcc\ falls less steeply than in the signal samples above $\mcc\sim 600$~\GeV. The minimum transverse mass $\mtcmin$, where $m_{\mathrm{T}}(c)=\sqrt{2p_{\mathrm{T}}(c) \times E_{\mathrm{T}}^{\mathrm{miss}}\times (1-\cos \Delta\phi(c, \vec{p}_{\mathrm{T}}^{\mathrm{miss}}))}$ is calculated for each $c$-tagged jet, is required to be at least 250~\GeV to reject \Wtaunujets{} backgrounds containing hadronically decaying $\tau$-leptons misidentified as $c$-tagged jets. Finally, various \metsig\ requirements are imposed to target signal models which are expected to have different kinematics. In addition to SR-HM1, SR-HM2, and SR-HM3, a signal region SR-HM-Disc with less stringent requirements is defined and is only used to set model-independent cross-section limits.

\begin{table}[htb]
\caption{Requirements of the SR-HM1--3 High-Mass regions, which are aimed at signal models with high top squark mass
($\gtrsim\qty{600}{\GeV}$), and SR-HM-Disc which is aimed at a broad range of models.}

% The next lines are included from the .//tex/SRAB.tex input file
\begin{center}
\begin{tabular}{l||c|c|c|c|} \hline\hline
\textbf{Variable} & \textbf{SR-HM1}&\textbf{SR-HM2}&\textbf{SR-HM3}&\textbf{SR-HM-Disc}\\ \hline \hline
Number of leptons & \multicolumn{4}{c|}{$=0$ } \\ \hline
Jet  $p_{\mathrm{T}}$ [GeV] & \multicolumn{4}{c|}{$> 40$ } \\ \hline
Number of $b$-tagged jets & \multicolumn{4}{c|}{$=0$ } \\ \hline
$p_{\mathrm{T}} (j_{1})$ [GeV] &  \multicolumn{4}{c|}{$> 250$ } \\ \hline
$\Delta\phi(\metvec,\vec{p}_{\mathrm{T}} (j_{4}))$ &  \multicolumn{4}{c|}{$> 0.4$ } \\ \hline
Number of $c$-tagged jets  & \multicolumn{4}{c|}{$\ge 2$ } \\ \hline
Leading jet is $c$-tagged & \multicolumn{4}{c|}{yes } \\ \hline

$m_{\mathrm{T}}^{\mathrm{min}}(c)$ [GeV]& \multicolumn{4}{c|}{$> 300$ } \\ \hline
$p_{\mathrm{T}} (j_{2})$ [GeV]& $> 200$ & \multicolumn{3}{c|}{$> 150$ } \\ \hline
\mcc\ [GeV]& $> 200$  & $ 200{-}400$  & $ 400{-}600$ & $>200$  \\ \hline
\metsig & ~~$> 22$  & \multicolumn{2}{c|}{$ 15{-}22$ } & ~~$>15$ \\ \hline
\hline
\end{tabular}
\end{center}

% End of text imported from the .//tex/SRAB.tex input file

\label{tab:srab_def}
\end{table}

\begin{table}[htb]
\caption{Requirements of the Compressed signal regions that target signal models with top-squark--neutralino mass differences between \qty{20}{\GeV} and \qty{175}{\GeV}. }
\def\arraystretch{1.2}

% The next lines are included from the .//tex/SRC.tex input file
\begin{center}
\begin{tabular}{l||c|c|c|c|} \hline\hline
\textbf{Variable} & \textbf{SR-Comp1}&\textbf{SR-Comp2}&\textbf{SR-Comp3}&\textbf{SR-Comp-1c}\\ \hline \hline
Number of leptons & \multicolumn{4}{c|}{$=0$ } \\ \hline
Jet  $p_{\mathrm{T}}$ [GeV] & \multicolumn{4}{c|}{$> 20$ } \\ \hline
Number of $b$-tagged jets & \multicolumn{4}{c|}{$=0$ } \\ \hline
$p_{\mathrm{T}} (j_{1})$ [GeV] &  \multicolumn{4}{c|}{$> 250$ } \\ \hline
$\Delta\phi(\metvec,\vec{p}_{\mathrm{T}} (j_{4}))$ &  \multicolumn{4}{c|}{$> 0.4$ } \\ \hline
Number of $c$-tagged jets & \multicolumn{3}{c|}{$\ge 2$ } & $\mathrm{=}~1$ \\ \hline
Leading jet is $c$-tagged & \multicolumn{4}{c|}{no } \\ \hline
$m_{\mathrm{T}}^{\mathrm{min}}(c)$ [GeV]& {--} & $> 80$ & $> 120$ & $< 250$  \\ \hline
\mcc\ [GeV]& \multicolumn{2}{c|}{--} & $> 100$  & {--} \\ \hline
$p^{\mathrm{CM}}_{\mathrm{T}}$ [GeV]& \multicolumn{3}{c|}{--} & ~~$< 10$  \\ \hline
$m_{\mathrm{V}}$ [GeV]& \multicolumn{3}{c|}{--} & ~~$< 80$  \\ \hline
$\Delta \phi(\metvec, j_{1}$) & \multicolumn{3}{c|}{--} & $> 2$  \\ \hline
\ncjs & \multicolumn{3}{c|}{$\ge 2$ } & $\mathrm{=}~1$  \\ \hline
$p^{\mathrm{ISR}}_{\mathrm{T}}$ [GeV]& $> 550$  & $> 500$  & $> 400$  & $> 700$  \\ \hline
$R_{\mathrm{ISR}}$ & $> 0.85$  & $ 0.75{-}0.85$  & $> 0.65$  & $ 0.9{-}1.0$  \\ \hline
\hline
\end{tabular}
\end{center}

% End of text imported from the .//tex/SRC.tex input file

\label{tab:rjsr_def}
\end{table}

Four Compressed signal regions, named SR-Comp1, SR-Comp2, SR-Comp3 and
SR-Comp-1c, are defined as shown in \cref{tab:rjsr_def}. All
Compressed signal regions are orthogonal to the High-Mass signal
regions as the leading jet is required to not be $c$-tagged. The most
compressed SUSY scenarios, with
$\Delta\sigPoint\approx\qty{20}{-}\qty{50}{\GeV}$, are targeted with
SR-Comp1 and SR-Comp-1c. SUSY scenarios with
$\Delta\sigPoint\approx\qty{80}{\GeV}$ are targeted with SR-Comp2, and
a looser, more general signal region, SR-Comp3, is designed to target
a wide range of larger mass splittings
$\qty{50}{\GeV}\lesssim\Delta\sigPoint\lesssim\qty{175}{\GeV}$. At
least two $c$-tagged jets are required for SR-Comp1, while only one is
required for SR-Comp-1c to increase signal contributions for the
low-\pt\ regime where the $c$-tagging efficiency decreases. SR-Comp-1c
also has a requirement of $\Delta \phi(\metvec, j_{1}) > 2$ on the
azimuthal angle separation between the leading jet and the \metvec.  In SR-Comp3, selections on \mtcmin{} and \mcc{} improve background rejection, with the latter serving to reduce background contributions from $g\to c\bar{c}$. There is some overlap between the Compressed signal regions: SR-Comp3 overlaps with SR-Comp1 and SR-Comp2, but SR-Comp1, SR-Comp2, and SR-Comp-1c are all orthogonal to one another, and SR-Comp-1c is orthogonal to SR-Comp3. Recursive Jigsaw Reconstruction~\cite{SUSY-2018-05} is used in the Compressed signal regions to enhance background rejection. In the presence of an ISR system, which consists of one or more jets produced by initial-state radiation, the $\stopone\stopone$ or $\scharmone\scharmone$ system is boosted in the transverse plane. Kinematic variables are then defined appropriately for the assignment of objects to either the ISR system or the sparticle (squarks plus LSPs) system. This method is equivalent to grouping the event objects according to the axis of maximum back-to-back \pt{} in the event's centre-of-mass (CM) frame, where the \pt{} of all objects sums vectorially to zero. The technique provides a suite of variables which are leveraged in the Compressed signal regions and include \risr{}, which is the projection of the invisible system's \pt{} direction vector onto the \myISR{} system's \pt{} direction vector. The ISR and sparticle systems tend to be approximately back-to-back in compressed SUSY events, resulting in \risr{} values
close to unity, while SM backgrounds tend to populate a wider range of \risr{} values. The vector sum of the jets' transverse momenta in the \myISR{} frame, \ptisr{}, is also used to reject background. For SM
backgrounds, \ptisr{} has a smoothly falling distribution, while for compressed SUSY signal models, a peak at higher \ptisr{} is seen, with the location of the peak typically being related
to the mass of the parent SUSY particle. The number of $c$-tagged jets assigned to the sparticle (\myS{}) frame of the ISR decay tree, \ncjs{}, is also utilized as a discriminating variable.
To suppress the increased SM contributions, SR-Comp-1c requires more stringent selections on \risr\ and \ptisr, and additional selections are also imposed on $p_{\myT}^{\myCM}$, the transverse momentum of the CM frame as evaluated in the laboratory frame, and $m_\myV$, the invariant mass of objects assigned to the visible (V) system.

Model-independent cross-section limits are set in the signal regions with the least stringent requirements (SR-HM1, SR-Comp3, and SR-HM-Disc) to search for any \enquote{beyond the Standard Model} (BSM) contributions. SR-HM-Disc has requirements similar to those in the other SR-HM regions, but it has a less stringent requirement on \mcc\ ($\mcc>200$~\GeV) and lacks the upper bound on \metsig\ applied in SR-HM2 and SR-HM3 (see \cref{tab:srab_def}).

% End of text imported from the .//tex/signal_regions.tex input file

% The next lines are included from the .//tex/backgrounds.tex input file
\subsection{Background estimation and the control regions}

\par The dominant backgrounds in all the signal regions are from processes with $W$ or $Z$ bosons produced in association with jets, with the bosons decaying as $W\to\ell\nu$ or $Z\to\nu\nu$ and producing real missing transverse momentum. The $Z$+jets background has similar contributions from $Z+c\bar{c}$, $Z+cj$, and $Z+jj$  (where $j$ indicates a non-charm quark) in all signal regions except SR-Comp-1c, where the $Z+jj$ background dominates. The $W$+jets background has larger contributions from the $W+cj$ and $W+jj$ components than from $W+c\bar{c}$. To reduce the systematic uncertainties, these backgrounds are estimated by extrapolating production-rate correction factors from the control regions (CRs) to the signal regions using the simulated samples. The extrapolation is performed in lepton multiplicity as illustrated in \cref{fig:bkg_strat} while other requirements in the CRs are made to be as close as possible to those in the signal regions. The normalization of the $Z$+jets and $W$+jets backgrounds is performed \emph{in situ} with the two-lepton (CRZ) and one-lepton (CRW) control regions, respectively. Three groups of CRs are designed for each background contribution: one set of CRs for the High-Mass signal regions, one set for the Compressed signal regions with two $c$-tagged jets, and one set for the Compressed signal region with one $c$-tagged jet. Contamination from signal processes was checked and found to be negligible in the defined CRs. Background processes that contribute less significantly are not normalized using control regions but using calculated cross-sections and include processes involving top quarks and multiple vector bosons. These processes are grouped together and referred to as \enquote{Other} processes in the remainder of this document.

\begin{figure}[htb]
\centering
\subfloat[]{\includegraphics[width=0.33\columnwidth]{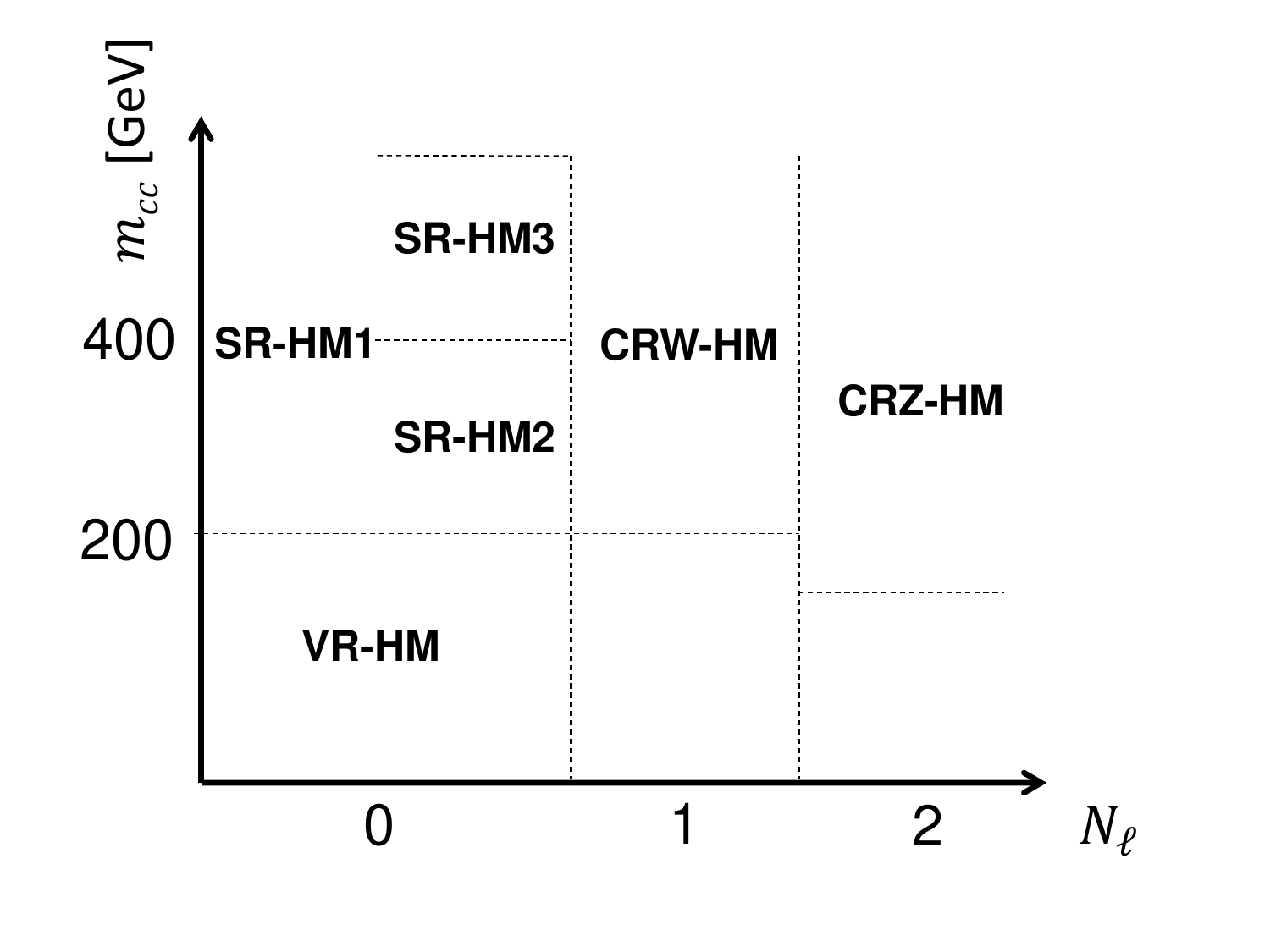}}
\subfloat[]{\includegraphics[width=0.33\columnwidth]{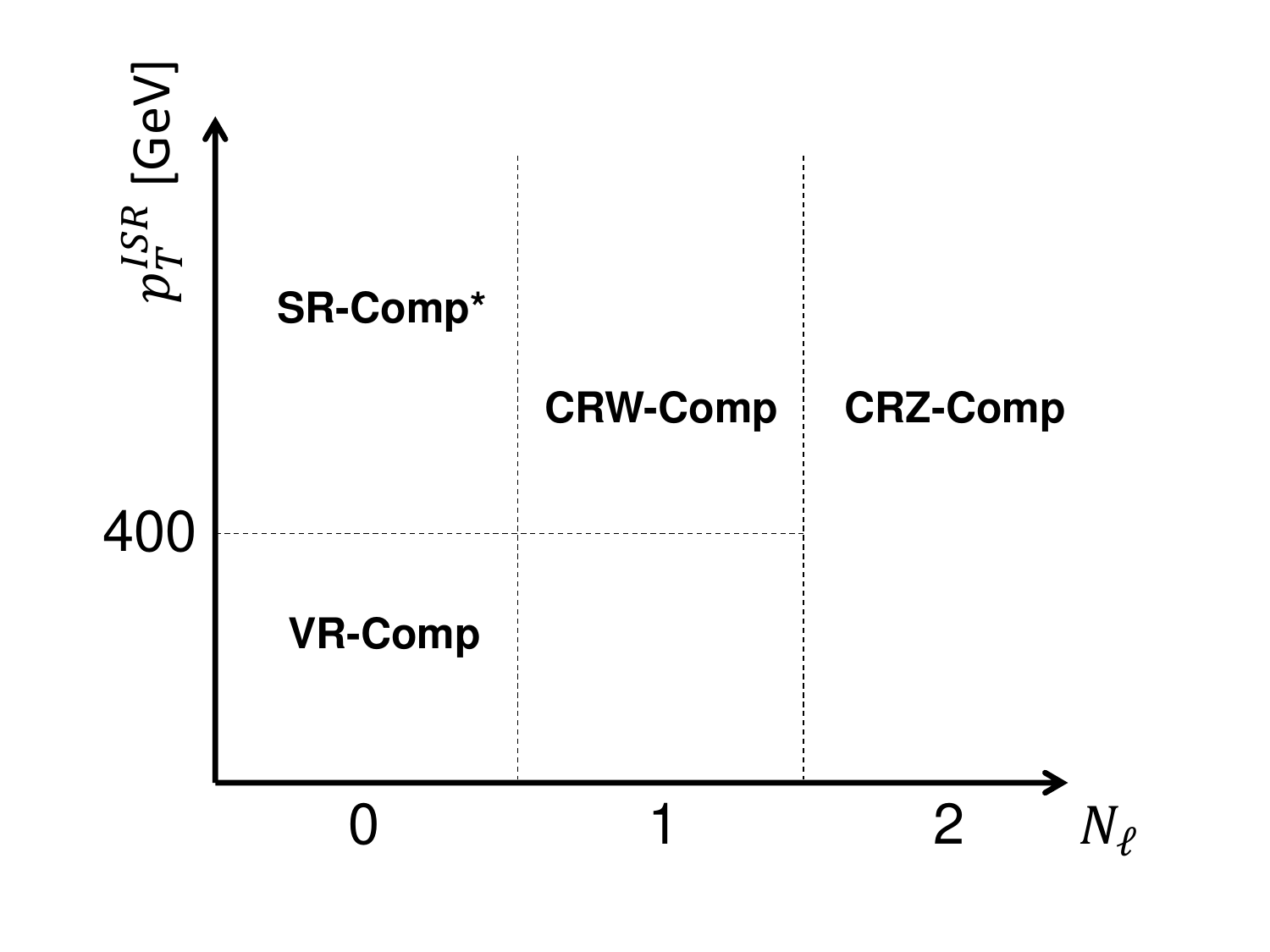}}
\subfloat[]{\includegraphics[width=0.33\columnwidth]{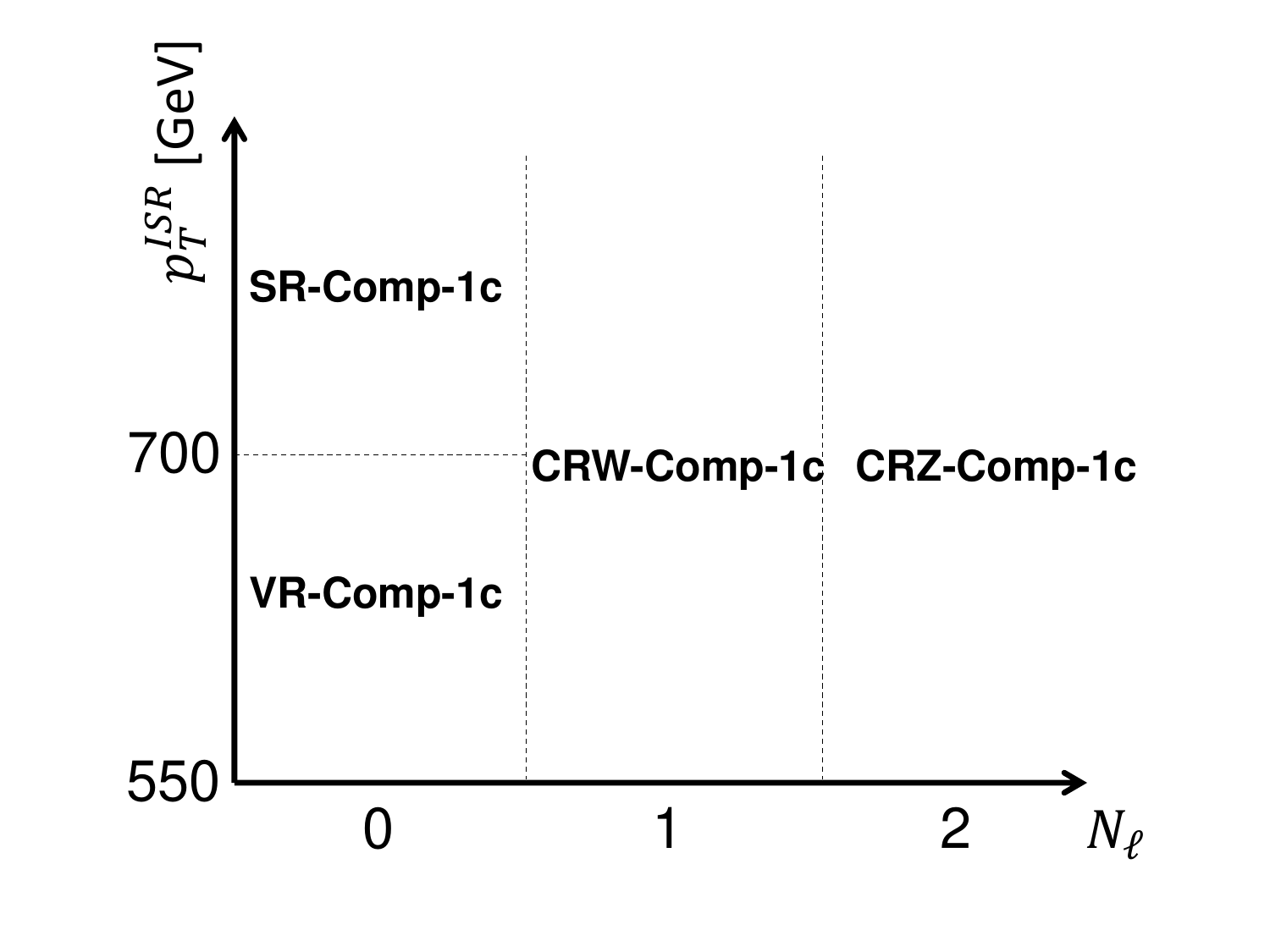}}
\caption{Phase-space partitioning between  the signal, control, and validation regions for the (a) High-Mass signal regions, (b) Compressed signal regions SR-Comp1, SR-Comp2 and SR-Comp3, and (c) SR-Comp-1c. Control
regions differ from the signal and validation regions by the electron and muon multiplicities (exactly one electron or muon in the $W$+jets control regions and two electrons or muons in the
$Z$+jets regions) while the validation regions veto events with leptons as in the signal regions. The asterisk in \enquote{SR-Comp*} is a wildcard character to denote all the Compressed signal regions which require two $c$-tagged jets.}
\label{fig:bkg_strat}
\end{figure}

\par The data for the $Z$+jets control regions (CRZ) were collected with the lowest unprescaled single-electron and single-muon triggers. These triggers have nearly constant efficiency for leptons with a \pt\ ($p_{\mathrm{T}} (\ell_{1})$) greater than 27~\GeV. To ensure a high-purity sample of $Z$+jets events, the invariant mass of the two signal leptons is required to be consistent with the mass of the $Z$ boson
($\qty{76}{\GeV}<m_{\ell\ell}<\qty{106}{\GeV}$). Further details of the CRZ selection are shown in \cref{tab:crz_def}.

\par For all and only the $Z$+jets control regions, \met\ and all the other variables that use it
(including \metsig) treat the leptons as invisible by subtracting their contribution from the \metvec\ computation; these variables are indicated with a
prime~($^\prime$). The treatment of the two leptons as invisible particles emulates the dominant contribution from $Z\rightarrow\nu\nu$ decays. To avoid significant extrapolation
uncertainties between signal and control regions related to jet flavor composition, the requirements on the number of $c$-tagged jets ($=1$ or $\ge2$) and on \mtcminprime\ are kept as
close as possible to those in the Compressed signal regions while also ensuring that the CRs provide sufficiently small statistical uncertainty in the $Z$+jets production rate. Distributions of selected kinematic
observables are shown in \cref{fig:crzDists}. These distributions illustrate the accuracy of the simulated samples' modelling of the data.

\begin{table}[htb]
\caption{Requirements of the $Z$+jets control regions for SR-HM1, SR-HM2, and SR-HM3 (CRZ-HM), SR-Comp1, SR-Comp2, and SR-Comp3 (CRZ-Comp), and SR-Comp-1c (CRZ-Comp-1c). Variables that are calculated using \met{} (including \metsig) use an \met{} variant that includes contributions from electrons/muons that are treated as invisible, and are indicated by a prime ($^\prime$).}
\def\arraystretch{1.3}

% The next lines are included from the .//tex/CRZ.tex input file
\begin{center}
\begin{tabular}{l||c|c|c|} \hline\hline
\textbf{Variable} & \textbf{CRZ-HM}&\textbf{CRZ-Comp}&\textbf{CRZ-Comp-1c}\\ \hline \hline
Used for & SR-HM &SR-Comp & SR-Comp-1c\\ \hline \hline

$p_{\mathrm{T}} (j_{1})$ & \multicolumn{2}{c|}{$> 250$ GeV} & $> 350$ GeV \\ \hline
Number of $c$-tagged jets & \multicolumn{2}{c|}{$\ge 2$}  & $\mathrm{=}~1$ \\ \hline
Leading jet is $c$-tagged & yes  & \multicolumn{2}{c|}{no } \\ \hline
Number of signal muons/electrons & \multicolumn{3}{c|}{$\mathrm{=}~2$ } \\ \hline
Passed single-electron/muon triggers & \multicolumn{3}{c|}{yes } \\ \hline
$p_{\mathrm{T}} (\ell_{1})$ [GeV]& \multicolumn{3}{c|}{~~$> 30$ } \\ \hline
$m_{\ell\ell}$ [GeV]& \multicolumn{3}{c|}{$ 76{-}106$ } \\ \hline
$E_{\mathrm{T}}^{\mathrm{miss}}$ [GeV]& \multicolumn{3}{c|}{$< 200$ } \\ \hline
$E_{\mathrm{T}}^{\mathrm{miss'}}$ [GeV]& \multicolumn{3}{c|}{$> 200$ } \\ \hline
\mcc\ [GeV]& $> 150$ & \multicolumn{2}{c|}{--} \\ \hline
$m_{\mathrm{T}}^{\mathrm{min'}}(c)$ [GeV]& $> 150$ & -- & $< 250$ \\ \hline
\metsigprime & ~~$> 10$  & \multicolumn{2}{c|}{--} \\ \hline
$N^{\mathrm{S'}}_{c\textrm{-jets}}$ & -- & $\ge 1$  & $\mathrm{=}~1$  \\ \hline
$R_{\mathrm{ISR}}'$ & -- & $> 0.75$  & $ 0.9{-}1.0$  \\ \hline
$p^{\mathrm{ISR'}}_{\mathrm{T}}$ [GeV]& \multicolumn{2}{c|}{--} & $> 550$  \\ \hline
$\Delta \phi(\metvecprime, j_{1}$) & \multicolumn{2}{c|}{--} & $> 2$  \\ \hline
\hline

\end{tabular}
\end{center}

% End of text imported from the .//tex/CRZ.tex input file

\label{tab:crz_def}
\end{table}

\begin{figure}[!htb]
\centering
\subfloat[]{\includegraphics[width=0.48\columnwidth]{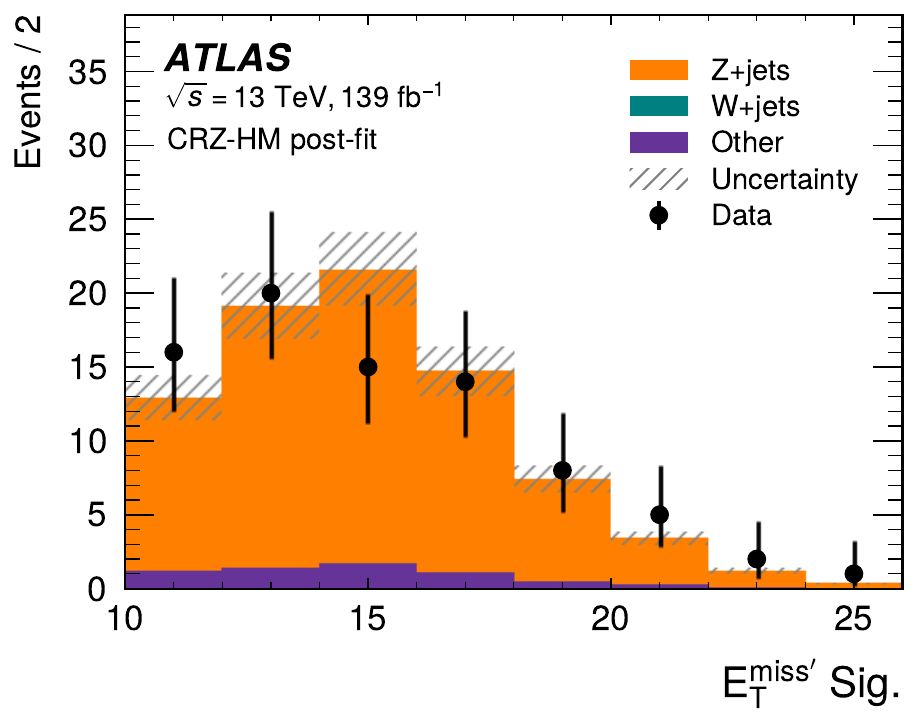}}
\subfloat[]{\includegraphics[width=0.48\columnwidth]{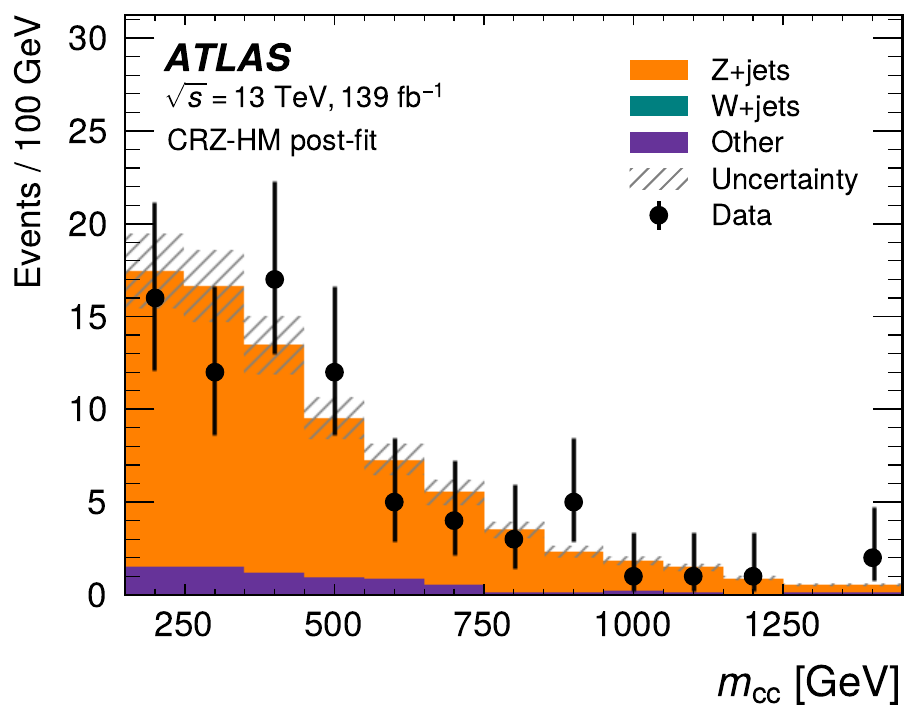}}\\
\subfloat[]{\includegraphics[width=0.48\columnwidth]{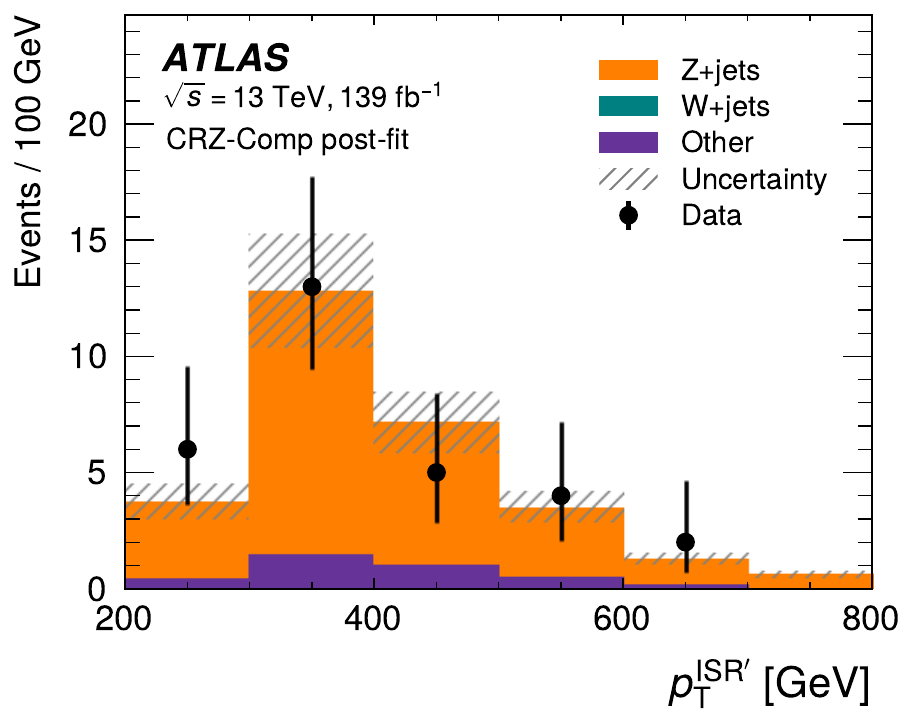}}
\subfloat[]{\includegraphics[width=0.48\columnwidth]{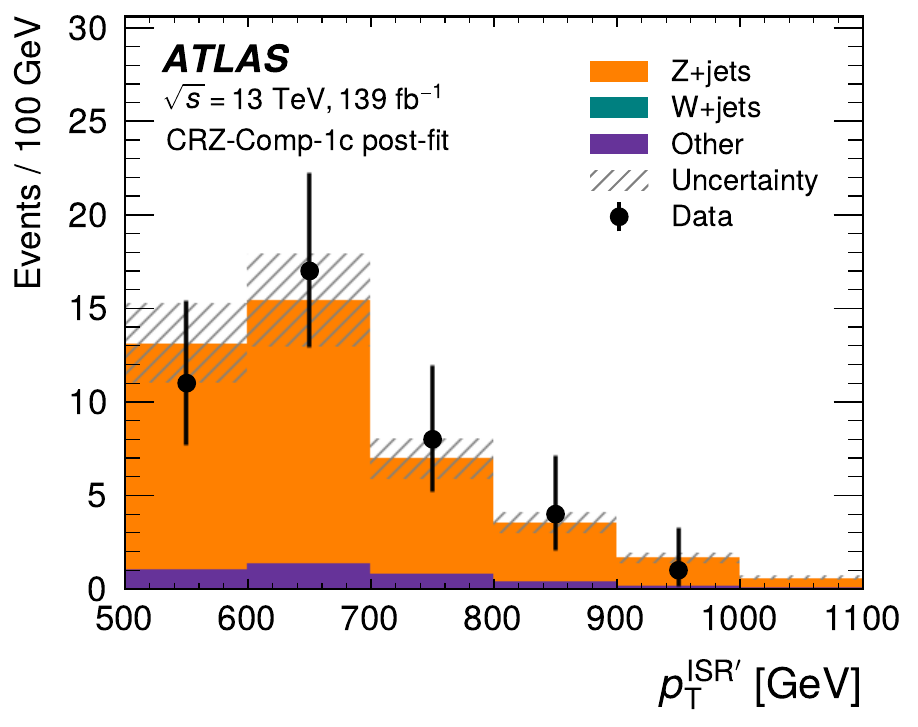}}
\caption{Distributions showing the level of agreement the between data (points) and the SM expectation (stacked histograms, after simultaneously fitting the $Z$+jets and $W$+jets backgrounds) in the $Z$+jets control regions for some
variables which have different selection criteria
in the signal regions: (a) \metsig\ in CRZ-HM, (b) \mcc\ in CRZ-HM, (c) \ptisrprime\ in CRZ-Comp, and (d) \ptisrprime\ in CRZ-Comp-1c. The hatched
uncertainty band around the total SM expectation
includes theory-based and detector-related systematic uncertainties and MC statistical uncertainties. Processes with top quarks and multiple
vector bosons are included in \enquote{Other}. The $W$+jets background contribution is
negligible and not visible in the plots. The right-most bin in each histogram does not
include the overflow entry but the $x$-axis range is chosen to include all observed data.}
\label{fig:crzDists}
\end{figure}

\par The $W$+jets control regions (CRW) require exactly one signal electron or muon. To keep the kinematic selection in the CRWs and the corresponding signal regions similar, all the CRWs require events to pass the \met{} trigger
and to have $\met > \qty{250}{\GeV}$. To ensure a high purity of $W$+jets background and to avoid contamination from $\ttbar$ processes, CRW-Comp includes a veto on events that include a $\tau$-lepton. CRW-Comp has a higher contamination from $\ttbar$ than is seen in CRW-HM because of the former's lower \mtcmin\ requirement, which is made to match the requirements in SR-Comp. CRW-Comp-1c contains events that have different kinematics due to its one-$c$-tag and high-\pt-ISR requirements, and thus does not require additional $\ttbar$ rejection from a $\tau$-lepton veto. A detailed description of all the CRW selections is shown in \cref{tab:crw_def}. \cref{fig:crwDists} shows the distributions of the main kinematic variables
used for extrapolation from the CRWs to the Compressed signal regions.

\begin{table}[htb]
\caption{Requirements of the $W$+jets control regions for SR-HM1, SR-HM2, and SR-HM3 (CRW-HM), SR-Comp1, SR-Comp2, and SR-Comp3 (CRW-Comp), and SR-Comp-1c (CRW-Comp-1c).}
\def\arraystretch{1.3}

% The next lines are included from the .//tex/CRW.tex input file
\begin{center}
\begin{tabular}{l||c|c|c|} \hline\hline
\textbf{Variable} & \textbf{CRW-HM}&\textbf{CRW-Comp}&\textbf{CRW-Comp-1c}\\ \hline \hline
Used for & SR-HM & SR-Comp & SR-Comp-1c\\ \hline \hline
$p_{\mathrm{T}} (j_{1})$ [GeV]& \multicolumn{2}{c|}{$> 250$} & $> 350$  \\ \hline
Number of $c$-tagged jets &\multicolumn{2}{c|}{$\ge 2$}  & $\mathrm{=}~1$  \\ \hline
Number of signal muons/electrons & \multicolumn{3}{c|}{$\mathrm{=}~1$ } \\ \hline
Passed $E_{\mathrm{T}}^{\mathrm{miss}}$ trigger & \multicolumn{3}{c|}{yes } \\ \hline
$E_{\mathrm{T}}^{\mathrm{miss}}$ [GeV]& \multicolumn{3}{c|}{$> 250$} \\ \hline
\mcc\ [GeV]& $> 200$ & \multicolumn{2}{c|}{--} \\ \hline
$m_{\mathrm{T}}^{\mathrm{min}}(c)$ [GeV]& $> 200$ & $> 120$ & $< 250$ \\ \hline
\metsig & $> 14$  & \multicolumn{2}{c|}{--} \\ \hline
Leading jet is $c$-tagged & yes  & \multicolumn{2}{c|}{no } \\ \hline
\ncjs\ & -- & $\ge 1$  & $\mathrm{=}~1$  \\ \hline
$p^{\mathrm{ISR}}_{\mathrm{T}}$ [GeV]& -- & $> 400$ & $> 550$ \\ \hline
$R_{\mathrm{ISR}}$ & -- & $> 0.65$  & $ 0.9{-}1.0$  \\ \hline
$\tau$ veto & -- & yes  & -- \\ \hline
$\Delta \phi(\metvec, j_{1}$) & \multicolumn{2}{c|}{--} & $> 2$  \\ \hline
\hline

\end{tabular}
\end{center}

% End of text imported from the .//tex/CRW.tex input file

\label{tab:crw_def}
\end{table}

\begin{figure}[htb]
\centering
\subfloat[]{\includegraphics[width=0.48\columnwidth]{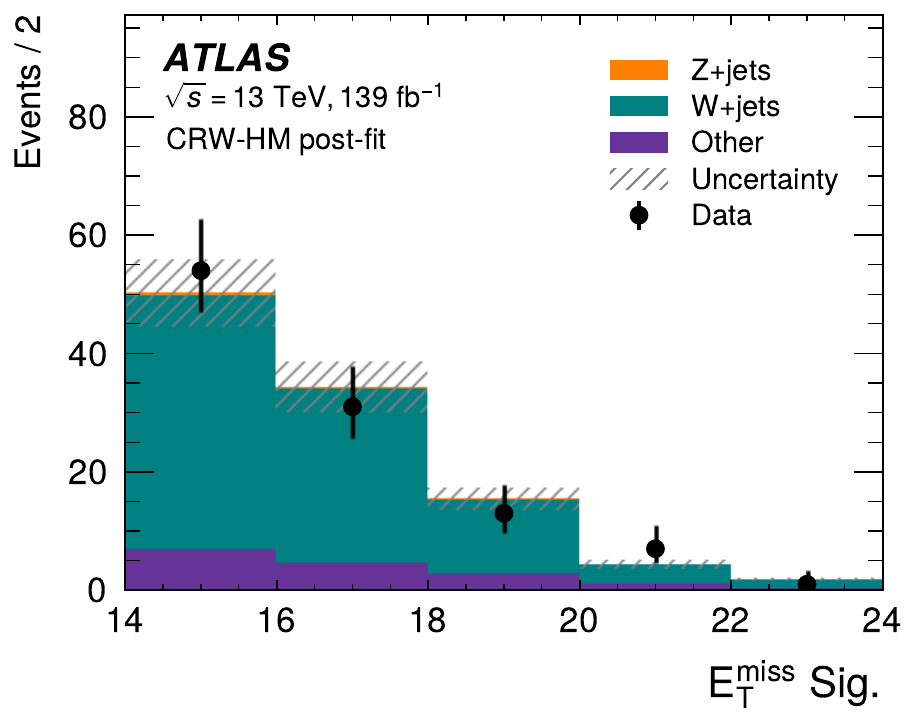} }
\subfloat[]{\includegraphics[width=0.48\columnwidth]{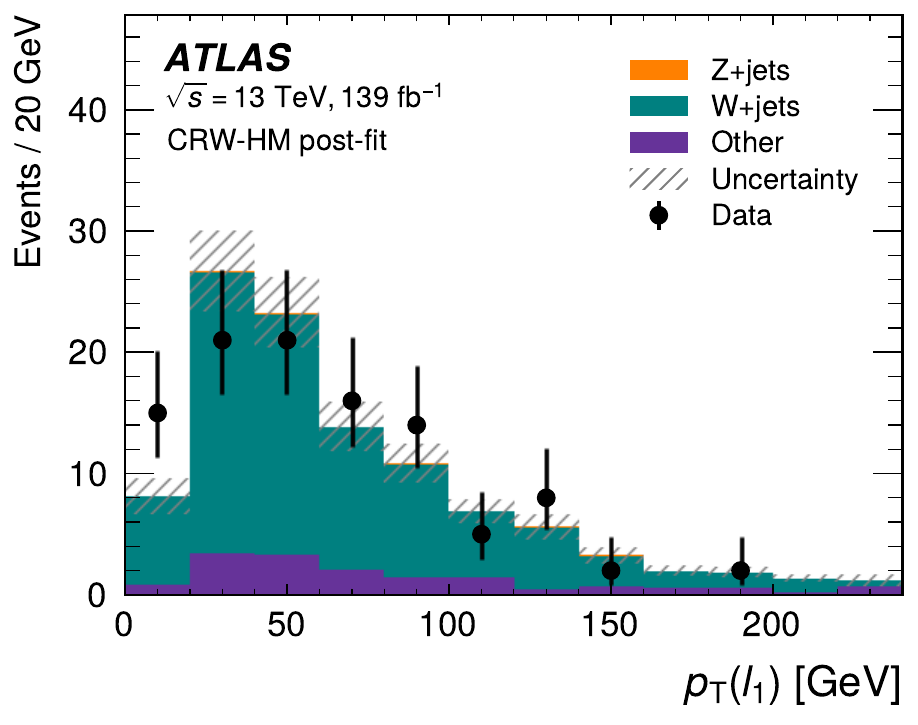}}\\
\subfloat[]{\includegraphics[width=0.48\columnwidth]{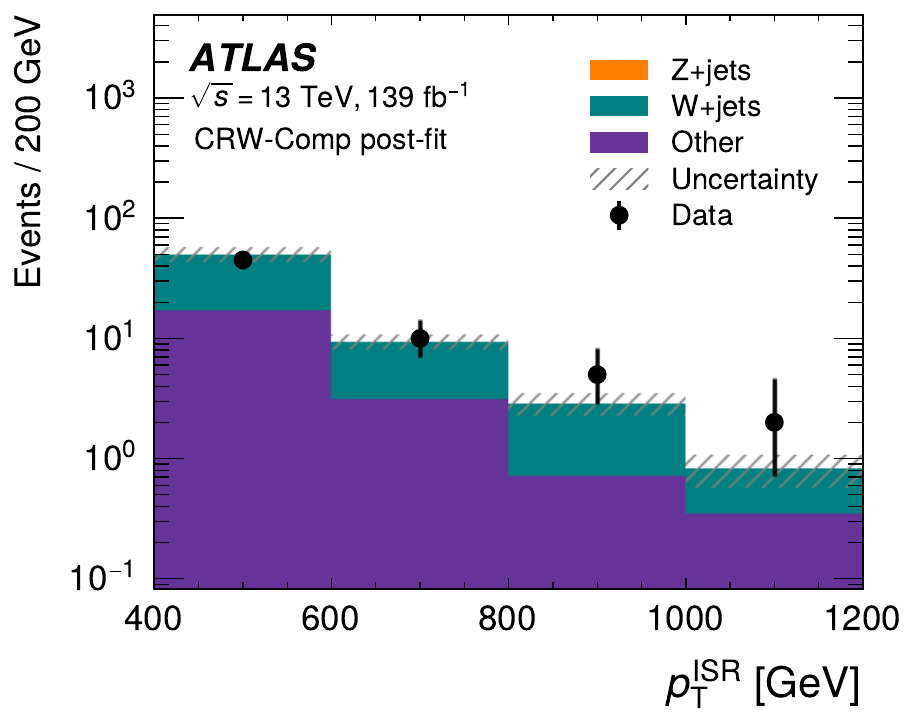}}
\subfloat[]{\includegraphics[width=0.48\columnwidth]{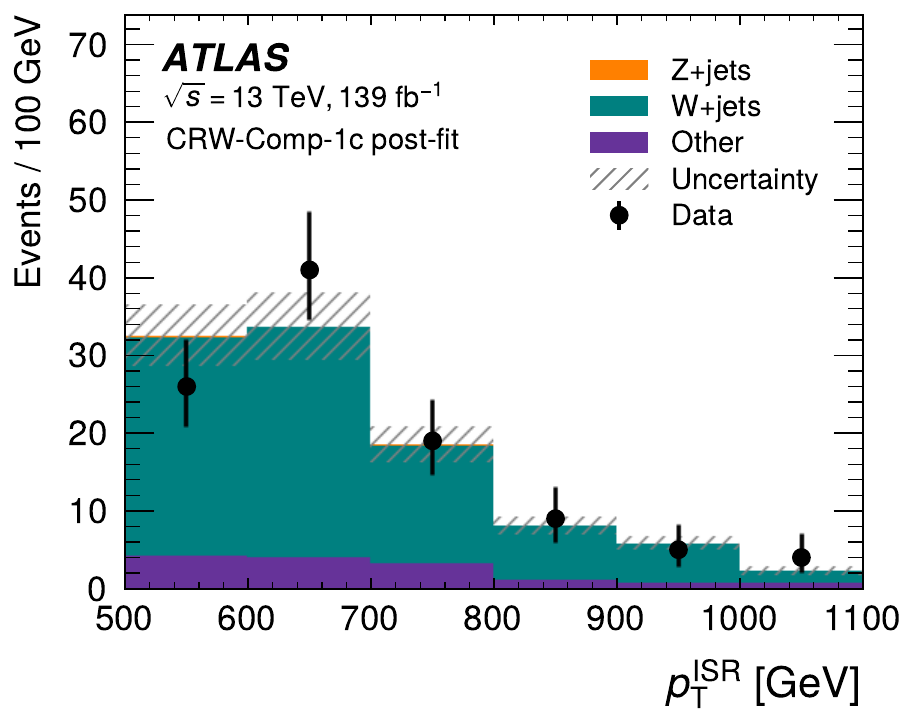}}

\caption{Distributions showing the level of agreement between the data (points) and the SM expectation (stacked histograms, after simultaneously
fitting the $Z$+jets and $W$+jets backgrounds) in the $W$+jets control regions for some variables which have different selection criteria in the
signal regions: (a) \metsig\ in CRW-HM, (b) \pt\ of leading lepton in CRW-HM, (c) \ptisr\ in CRW-Comp, and (d) \ptisr\ in CRW-Comp-1c.
The hatched uncertainty band around the total SM expectation
includes theory-based and detector-related systematic uncertainties and MC statistical uncertainties.
Processes with top quarks and multiple vector bosons are  included in
\enquote{Other}. The $Z$+jets background contribution is
negligible and not visible in the plots. The right-most bin in each histogram does not
include the overflow entry but the $x$-axis range is chosen to include all observed data.}

\label{fig:crwDists}
\end{figure}

\subsection{Validation of background estimates}

\par The validation regions are used to test the accuracy of the procedure used to extrapolate the background from the control region to the
signal regions before unblinding data in the signal regions.
The parameter used in the extrapolation between the control and signal regions is the electron and muon multiplicity. Therefore, the validation regions veto events with leptons as do the signal regions. The
validation regions are designed to have a background composition that consists of the dominant background processes expected in the signal regions, i.e.\ $Z$+jets and $W$+jets.
The validation regions are orthogonal to the signal regions and control regions and have sufficient event yields to test the extrapolation procedure with reasonable accuracy. The validation regions are also
designed to avoid large contamination from any signal model considered in this search, with less than 20\% contamination seen for the models not excluded by the previous ATLAS search.

\par To validate the extrapolation of the backgrounds from the control regions CRW-HM and CRZ-HM to the High-Mass signal regions SR-HM1, SR-HM2 and SR-HM3, a single validation region VR-HM is used, with an expected $Z$+jets and $W$+jets contribution of 45\% and 26\%, respectively. The
requirements for VR-HM are the same as in the signal region, except for changes which reduce the signal contamination such as a reversal of the \mcc\ requirement, an absence of \metsig\ and $p_{\mathrm{T}} (j_{2})$ requirements, and a
less stringent requirement of $\mtcmin>200$~\GeV{} (as compared to $\mtcmin>300$~\GeV\ in the signal regions).

\par The normalization procedure for Compressed regions that require at least two charm-tagged jets, i.e.\ SR-Comp1, SR-Comp2, and SR-Comp3, is validated in one region, VR-Comp. This region is
orthogonal to the signal regions and has low signal contamination because it has a reversed \ptisr\ requirement ($\ptisr<400$ GeV). The other requirements, including the veto on events with leading $c$-tagged jets
that makes the Compressed regions orthogonal to the High-Mass regions, are kept the same as in the signal regions, with the following four exceptions. The $c$-tagged jet multiplicity in the sparticle
frame, \ncjs, is loosened to be greater than or equal to one (rather than two in the signal regions). The requirement on \risr\ is also less stringent in VR-Comp
($\risr>0.65$) than in SR-Comp1 ($\risr>0.85$) and SR-Comp2 ($0.75<\risr<0.85$) but is the same as in SR-Comp3. Additionally, \mtcmin\ is required to be greater than 80~\GeV{} in VR-Comp, whereas there is no
requirement on \mtcmin\ in SR-Comp1. SR-Comp2 also requires a minimum of 80~\GeV, and SR-Comp3 requires $\mtcmin>120$ GeV. Finally, no requirement is made on \mcc\ in VR-Comp, as is the case in SR-Comp1
and SR-Comp2, but which differs from the $\mcc>100$~\GeV{} requirement in SR-Comp3.

\par The one-$c$-tag validation region is identical to SR-Comp-1c with the exception of the defining reversal of the \ptisr\ requirement relative to the signal region and the widening of the \risr\ window from $0.9<\risr<1.0$ in SR-Comp-1c to $0.85<\risr<1.05$. The full set of selections for all validation regions is shown in \cref{tab:vr_def}, while distributions of the main kinematic variables that differ between the validation regions and the signal regions are shown in \cref{fig:vrDists}.

\begin{table}[htb]
\caption{Requirements of the validation regions.}
\def\arraystretch{1.3}

% The next lines are included from the .//tex/VR.tex input file
\begin{center}
\begin{tabular}{l||c|c|c|} \hline\hline
\textbf{Variable} & \textbf{VR-HM}&\textbf{VR-Comp}&\textbf{VR-Comp-1c}\\ \hline \hline
Validates backgrounds in & SR-HM & SR-Comp & SR-Comp-1c\\ \hline \hline
Number of $c$-tagged jets &\multicolumn{2}{c|}{$\ge 2$}  & $\mathrm{=}~1$  \\ \hline
Passed $E_{\mathrm{T}}^{\mathrm{miss}}$ trigger & \multicolumn{3}{c|}{yes } \\ \hline
$E_{\mathrm{T}}^{\mathrm{miss}}$ [GeV]& \multicolumn{3}{c|}{$> 250$} \\ \hline
\mcc\ [GeV]& $> 200$ & \multicolumn{2}{c|}{--} \\ \hline
$m_{\mathrm{T}}^{\mathrm{min}}(c)$ [GeV]& $> 200$ & $> 80$ & $>100$ \\ \hline
Leading jet is $c$-tagged & yes  & \multicolumn{2}{c|}{no } \\ \hline
\ncjs\ & -- & $\ge 1$  & $\mathrm{=}~1$  \\ \hline
$p^{\mathrm{ISR}}_{\mathrm{T}}$ [GeV]& -- & \multicolumn{2}{c|}{$< 400$} \\ \hline
$R_{\mathrm{ISR}}$ & -- & $> 0.65$  & $ 0.85{-}1.05$  \\ \hline
$\tau$ veto & \multicolumn{3}{c|}{yes} \\ \hline
$\Delta \phi(\metvec, j_{1}$) & \multicolumn{2}{c|}{--} & ~~$> 2$  \\ \hline
$p^{\mathrm{CM}}_{\mathrm{T}}$ [GeV]& \multicolumn{2}{c|}{--} & $< 10$  \\ \hline
$m_{\mathrm{V}}$ [GeV]& \multicolumn{2}{c|}{--} & $< 80$  \\ \hline
\hline
\end{tabular}
\end{center}

% End of text imported from the .//tex/VR.tex input file

\label{tab:vr_def}
\end{table}

\begin{figure}[htb]
\centering
\subfloat[]{\includegraphics[width=0.48\columnwidth]{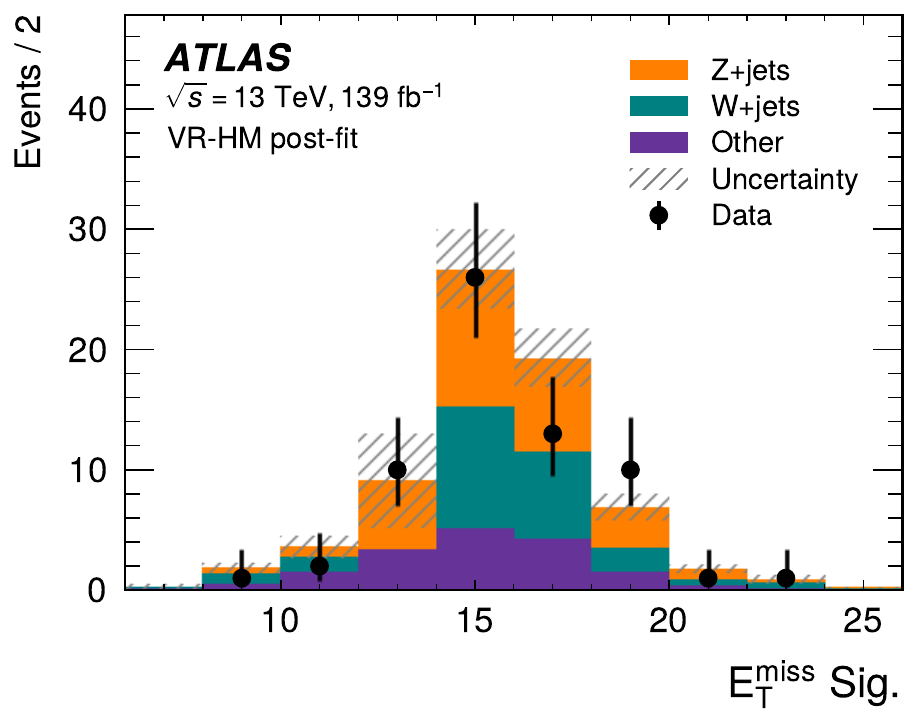}}
\subfloat[]{\includegraphics[width=0.48\columnwidth]{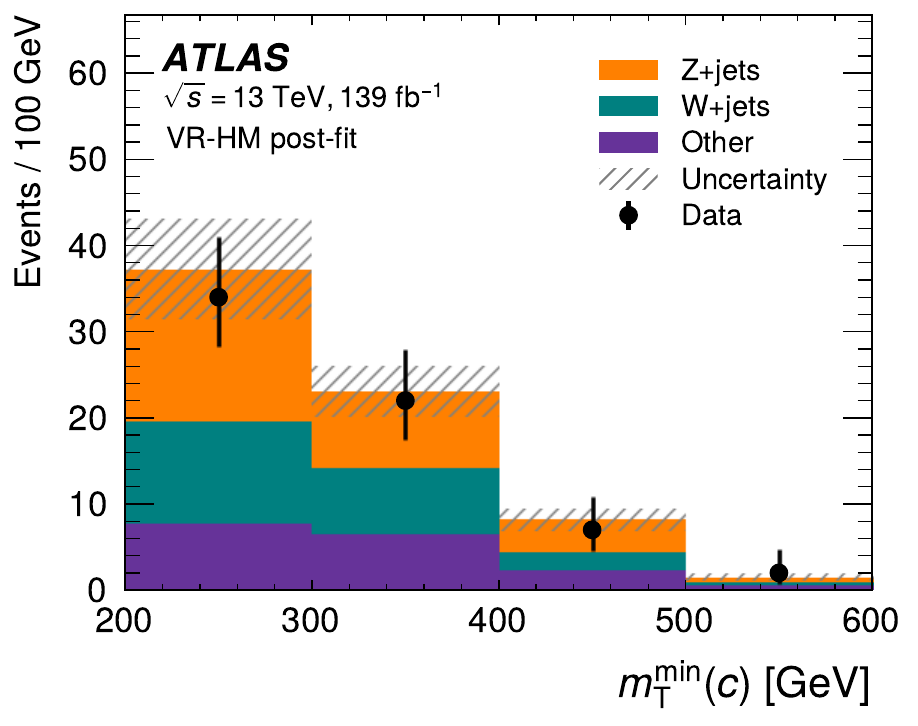}}\\
\subfloat[]{\includegraphics[width=0.48\columnwidth]{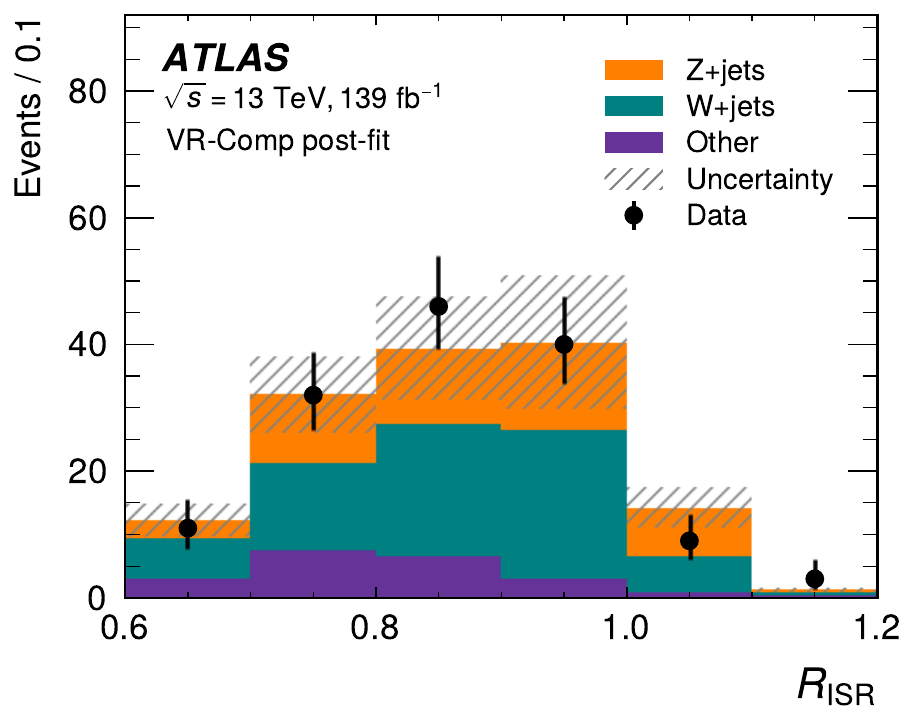}}
\subfloat[]{\includegraphics[width=0.48\columnwidth]{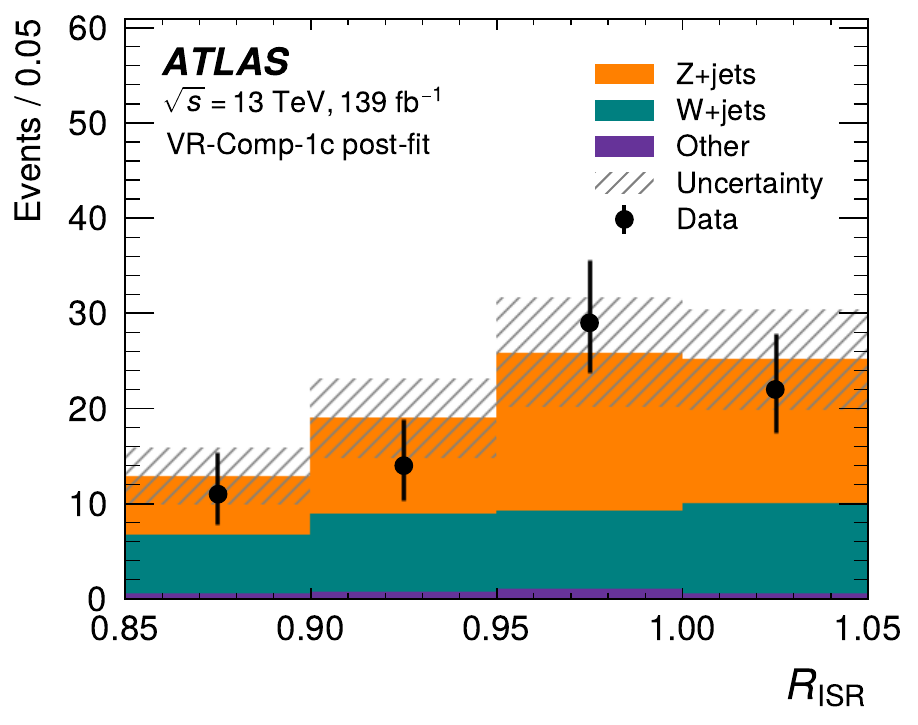}}
\caption{Distributions, in various validation regions, showing the level of agreement between the data (points) and the SM expectation (stacked histograms,
after simultaneously fitting the $Z$+jets and $W$+jets backgrounds) for variables which have different selection criteria in the signal regions:
(a) \metsig\ in VR-HM, (b) \mtcmin\ in VR-HM, (c) \risr\ in VR-Comp, and (d) \risr\ in VR-Comp-1c.
The hatched uncertainty band around the total SM expectation
includes theory-based and detector-related systematic uncertainties
and MC statistical uncertainties. Processes with top quarks and
multiple vector bosons are included in \enquote{Other}. The right-most bin in each histogram does not
include the overflow entry but the $x$-axis range is chosen to include all observed data.}
\label{fig:vrDists}
\end{figure}

% End of text imported from the .//tex/backgrounds.tex input file

% The next lines are included from the .//tex/systematics.tex input file
\section{Systematic uncertainties}

\par The effect of systematic uncertainties is taken into account for the limited accuracy of the background and signal predictions. To properly treat
correlations of uncertainties across the signal and control regions, the uncertainties are separated by source. There are uncertainties for the simulated detector response, integrated luminosity~\cite{ATLAS-CONF-2019-021}, number of simulated events, predicted cross-sections,
parton distribution functions, underlying event and minimum-bias $pp$ collisions, parton hadronization models,
limited accuracy of matrix element and parton shower calculations, and experimental effects. Below is a discussion of the most relevant sources of uncertainty affecting the search.

\par The uncertainty in the measured $c$-tagging efficiency is one of the dominant sources of uncertainty because all the signal and control regions require events with one or more
$c$-tagged jets. The simulated event samples are reweighted so that the predicted and measured identification efficiencies of $c$-tagged jets are identical and so the uncertainties are
propagated. The uncertainty in the $c$-tagging efficiency also depends on the jet \pt\ and whether the tagged jet is initiated by a $b$-quark, $c$-quark, or light
quark~\cite{FTAG-2020-08,FTAG-2018-01,FTAG-2019-02}.

\par The uncertainties in the predicted jet energies are driven by the accuracy of the measured jet energies and resolution~\cite{JETM-2018-05}. There is an uncertainty in the
contribution from jets that originate from the overlapping $pp$ interactions but are mistakenly counted as coming from the hard scatter~\cite{PERF-2014-03}. This
includes the efficiency of identifying jets originating from secondary $pp$ interactions and the accuracy of simulated minimum-bias interactions.

\par The theoretical uncertainties for the $W$+jets and $Z$+jets backgrounds are estimated with \SHERPA. Uncertainties due to inaccuracies in the matrix element
calculation and parton showering are estimated by varying the generator's parameters. The envelope of the associated predicted distributions is taken as the uncertainty. The matrix element
matching scale is varied between 15~\GeV and 30~\GeV. Its nominal value is 20~\GeV. The renormalization and factorization scales are varied by factors of 0.5 and 2 independently. The impact of
the PDF uncertainties is taken into account in addition to the generator uncertainties.

\par The secondary backgrounds such as top-quark pairs produced in association with jets~\cite{TOPQ-2017-12} and vector-boson pairs produced in association with jets, labeled as
\enquote{Other} in the figures and tables, have a minor effect on the total background estimate. A normalization uncertainty of 25\% is assigned and treated as fully correlated across
regions. This uncertainty estimate, is meant to cover cross-section normalization and modeling uncertainties and does not contribute significantly to the total uncertainty. Treating the
uncertainty uncorrelated between regions was found to increase the total expected background contribution by at most one standard deviation in SR-Comp2 and SR-Comp3 with a negligible effect
on the final result.

\par Additionally, for signal predictions (i.e.\ SUSY and LQ), theoretical uncertainties are taken as an envelope of the results obtained when varying the factorization and renormalization scales independently by factors of 0.5 and 2. The Var3c parameter (which corresponds to the strong coupling parameter, \alphas, used for ISR) of the A14 underlying-event tune is also varied. The contribution from these uncertainties is largest in the compressed regions, where it reaches 20\%.

% End of text imported from the .//tex/systematics.tex input file

% The next lines are included from the .//tex/results_conclusion.tex input file
\clearpage
\section{Results and interpretation}
\label{sec:result}

\par A statistical analysis of the data is performed with a simultaneous likelihood fit~\cite{Cowan:2010js,cabinetry,pyhf_joss} to the number of signal and background events in the control and signal regions; effects of systematic uncertainties are included through nuisance parameters that change the expected contributions across regions in a correlated manner.
The likelihood probability is calculated by combining Poisson probabilities for the signal, validation, and control regions with a Gaussian probability distribution for each systematic uncertainty. The $W$+jets and $Z$+jets background normalizations are free-floating in the fit, with three normalization factors for each (i.e.\ for High-Mass, Compressed, and one-$c$-tag regions).

\par A \enquote{background-only} fit is performed for the High-Mass (CRZ-HM, CRW-HM), Compressed (CRZ-Comp, CRW-Comp), and one-$c$-tag (CRZ-Comp-1c, CRW-Comp-1c) regions using the likelihood in the control regions and extrapolating the normalization of $Z$+jets and $W$+jets
contributions into the validation or signal regions (thus not considering the data yield in the signal and validation regions). The fitting strategy reduces systematic uncertainties in the signal regions because the control regions are chosen to minimize
the extrapolation uncertainties. Thus the effect on the $W$+jets and $Z$+jets background estimates due to systematic uncertainties is similar in the control and signal regions and partially cancels out. This strategy is especially effective in the High-Mass regions, where the total uncertainty of the SM backgrounds is reduced from 30\%--37\% to 13\%--17\%. The systematic uncertainties that contribute the most to the total post-fit uncertainty are related to $c$-tagging and $V$+jets theoretical uncertainties. Theoretical systematic uncertainties are dominated by the variations of the scales, with contributions to the total uncertainty ranging from 14\% to 36\%,
while the largest contributions from $c$-tagging-related systematic uncertainties to the total uncertainty of the background estimate in the signal regions range from 8\% to 14\%.

\par No significant excesses over the expected SM backgrounds in the validation and signal regions are observed when considering all statistical and systematic (experimental and theory) uncertainties, as shown in \cref{tab:high_fit_results,tab:compressedAB_fit_results} and \cref{fig:srVRSummary}. The extracted normalization factors for the $W$+jets and $Z$+jets backgrounds are shown
in~\cref{tab:norm} and are generally consistent with unity given their associated uncertainties, with the normalization factors for $W$+jets background being larger across all control regions. Selected kinematic distributions in the signal regions are shown in \cref{fig:srDists}.

\begin{DIFnomarkup}
\begin{table}[htb]
\centering
\caption{Post-fit event yields for the High-Mass control,
validation, and signal regions. The $Z$+jets and $W$+jets
backgrounds are normalized by fits within the control
regions. Processes with top quarks and multiple vector bosons
are included in \enquote{Other}. Pre-fit event yields for a
representative signal point for each of the signal regions are also shown.}
\small
\label{tab:high_fit_results}
\makebox[\textwidth]{

% The next lines are included from the .//tex/CRABYieldTable.tex input file
\begin{tabular}{
S[table-alignment=left]
S[table-number-alignment=left]
S[table-number-alignment=left]
S[table-number-alignment=left]
S[table-number-alignment=left]
S[table-number-alignment=left]
S[table-number-alignment=left]
}
\hline\hline
& \multicolumn{2}{c}{CRW-HM} & \multicolumn{2}{c}{CRZ-HM} & \multicolumn{2}{c}{VR-HM}\\ \hline{$Z$+jets} & \multicolumn{2}{l}{$\phantom{00}0.66 \pm \phantom{0}0.12$} & \multicolumn{2}{l}{$74\phantom{.000} \pm 9$} & \multicolumn{2}{l}{$31 \pm 5$}\\
{$W$+jets} & \multicolumn{2}{l}{$\phantom{0}90\phantom{.00} \pm 11$} & \multicolumn{2}{l}{$\phantom{0}0.030 \pm 0.009$} & \multicolumn{2}{l}{$22 \pm 4$}\\
{Other} & \multicolumn{2}{l}{$\phantom{0}15\phantom{.00} \pm \phantom{0}4$} & \multicolumn{2}{l}{$\phantom{0}6.6\phantom{00} \pm 1.8$} & \multicolumn{2}{l}{$17 \pm 5$}\\
\hline{Total SM} & \multicolumn{2}{l}{$106\phantom{.00} \pm 10$} & \multicolumn{2}{l}{$81\phantom{.000} \pm 9$} & \multicolumn{2}{l}{$70 \pm 8$}\\ \hline{Observed} & \multicolumn{1}{l}{$106$} &  & \multicolumn{1}{l}{$81$} &  & \multicolumn{1}{l}{$65$} & \\
\hline\hline
\end{tabular}

% End of text imported from the .//tex/CRABYieldTable.tex input file

}
\newline
\vspace*{0.1cm}
\newline
\def\arraystretch{1.5}

% The next lines are included from the .//tex/SRABYieldTable.tex input file
\begin{tabular}{
S[table-alignment=left]
S[table-number-alignment=left]
S[table-number-alignment=left]
S[table-number-alignment=left]
S[table-number-alignment=left]
S[table-number-alignment=left]
S[table-number-alignment=left]
S[table-number-alignment=left]
S[table-number-alignment=left]
}
\hline\hline
& \multicolumn{2}{c}{SR-HM1} & \multicolumn{2}{c}{SR-HM2} & \multicolumn{2}{c}{SR-HM3} & \multicolumn{2}{c}{SR-HM-Disc}\\ \hline{$Z$+jets} & \multicolumn{2}{l}{$\phantom{0}5.2\phantom{0} \pm 1.3$} & \multicolumn{2}{l}{$12.3 \pm 3.4$} & \multicolumn{2}{l}{$14.7 \pm 2.7$} & \multicolumn{2}{l}{$\phantom{0}58\phantom{.0} \pm \phantom{0}9$}\\
{$W$+jets} & \multicolumn{2}{l}{$\phantom{0}3.2\phantom{0} \pm 0.7$} & \multicolumn{2}{l}{$\phantom{0}9.8 \pm 1.7$} & \multicolumn{2}{l}{$12.5 \pm 2.1$} & \multicolumn{2}{l}{$\phantom{0}43\phantom{.0} \pm \phantom{0}6$}\\
{Other} & \multicolumn{2}{l}{$\phantom{0}0.88 \pm 0.33$} & \multicolumn{2}{l}{$\phantom{0}1.6 \pm 0.6$} & \multicolumn{2}{l}{$\phantom{0}2.2 \pm 0.7$} & \multicolumn{2}{l}{$\phantom{00}9.0 \pm \phantom{0}2.5$}\\
\hline{Total SM} & \multicolumn{2}{l}{$\phantom{0}9.3\phantom{0} \pm 1.6$} & \multicolumn{2}{l}{$24 \pm 4$} & \multicolumn{2}{l}{$29 \pm 4$} & \multicolumn{2}{l}{$110\phantom{.0} \pm 12$}\\ \hline{Observed} & \multicolumn{1}{l}{$13$} &  & \multicolumn{1}{l}{$31$} &  & \multicolumn{1}{l}{$27$} &  & \multicolumn{1}{l}{$133$} & \\
\hline{$\sigPointNoC$=$(1000,1)\,\GeV$} & \multicolumn{2}{l}{$\phantom{0}6.6\phantom{0} \pm 1.0$} & \multicolumn{1}{l}{} &  & \multicolumn{1}{l}{} &  & \multicolumn{2}{l}{$\phantom{0}12.7 \pm \phantom{0}2.1$}\\
\hline{$\sigPointNoC$=$(750,450)\,\GeV$} & \multicolumn{1}{l}{} &  & \multicolumn{2}{l}{$17.3 \pm 2.6$} & \multicolumn{2}{l}{$12.6 \pm 2.1$} & \multicolumn{2}{l}{$\phantom{0}41\phantom{.0} \pm \phantom{0}7$}\\
\hline\hline
\end{tabular}

% End of text imported from the .//tex/SRABYieldTable.tex input file

\end{table}
\end{DIFnomarkup}

\begin{DIFnomarkup}
\begin{table}[htb]
\centering
\caption{Post-fit event yields for the Compressed control, validation, and signal regions. The $Z$+jets and $W$+jets backgrounds are normalized by fits within the control regions. Processes with top quarks and multiple vector bosons are included in \enquote{Other}. Pre-fit event yields for a representative signal point in each signal region are also shown.}
\small
\label{tab:compressedAB_fit_results}
\makebox[\textwidth]{

% The next lines are included from the .//tex/CRCYieldTable.tex input file
\begin{tabular}{
S[table-alignment=left]
S[table-number-alignment=left]
S[table-number-alignment=left]
S[table-number-alignment=left]
S[table-number-alignment=left]
S[table-number-alignment=left]
S[table-number-alignment=left]
S[table-number-alignment=left]
S[table-number-alignment=left]
S[table-number-alignment=left]
S[table-number-alignment=left]
S[table-number-alignment=left]
S[table-number-alignment=left]
}
\hline\hline
& \multicolumn{2}{c}{CRW-Comp} & \multicolumn{2}{c}{CRZ-Comp} & \multicolumn{2}{c}{CRW-Comp-1c} & \multicolumn{2}{c}{CRZ-Comp-1c} & \multicolumn{2}{c}{VR-Comp} & \multicolumn{2}{c}{VR-Comp-1c}\\ \hline{$Z$+jets} & \multicolumn{2}{l}{$\phantom{0}0.33 \pm \phantom{0}0.12$} & \multicolumn{2}{l}{$26\phantom{.0} \pm 6$} & \multicolumn{2}{l}{$\phantom{00}0.65 \pm \phantom{0}0.18$} & \multicolumn{2}{l}{$38\phantom{.0} \pm 7$} & \multicolumn{2}{l}{$\phantom{0}48 \pm 12$} & \multicolumn{2}{l}{$48\phantom{.0} \pm 13$}\\
{$W$+jets} & \multicolumn{2}{l}{$41\phantom{.00} \pm 10$} & \multicolumn{2}{c}{--} & \multicolumn{2}{l}{$\phantom{0}89\phantom{.00} \pm 11$} & \multicolumn{2}{c}{--} & \multicolumn{2}{l}{$\phantom{0}71 \pm 25$} & \multicolumn{2}{l}{$32\phantom{.0} \pm 11$}\\
{Other} & \multicolumn{2}{l}{$22\phantom{.00} \pm \phantom{0}6$} & \multicolumn{2}{l}{$\phantom{0}3.9 \pm 1.2$} & \multicolumn{2}{l}{$\phantom{0}15\phantom{.00} \pm \phantom{0}4$} & \multicolumn{2}{l}{$\phantom{0}4.0 \pm 1.1$} & \multicolumn{2}{l}{$\phantom{0}21 \pm \phantom{0}6$} & \multicolumn{2}{l}{$\phantom{0}2.8 \pm \phantom{0}0.8$}\\
\hline{Total SM} & \multicolumn{2}{l}{$63\phantom{.00} \pm \phantom{0}8$} & \multicolumn{2}{l}{$30\phantom{.0} \pm 5$} & \multicolumn{2}{l}{$105\phantom{.00} \pm 10$} & \multicolumn{2}{l}{$42\phantom{.0} \pm 6$} & \multicolumn{2}{l}{$139 \pm 27$} & \multicolumn{2}{l}{$83\phantom{.0} \pm 17$}\\ \hline{Observed} & \multicolumn{1}{l}{$63$} &  & \multicolumn{1}{l}{$30$} &  & \multicolumn{1}{l}{$105$} &  & \multicolumn{1}{l}{$42$} &  & \multicolumn{1}{l}{$141$} &  & \multicolumn{1}{l}{$76$} & \\
\hline\hline
\end{tabular}

% End of text imported from the .//tex/CRCYieldTable.tex input file

}
\newline
\vspace*{0.1cm}
\newline
\def\arraystretch{1.5}

% The next lines are included from the .//tex/SRCYieldTable.tex input file
\begin{tabular}{
S[table-alignment=left]
S[table-number-alignment=left]
S[table-number-alignment=left]
S[table-number-alignment=left]
S[table-number-alignment=left]
S[table-number-alignment=left]
S[table-number-alignment=left]
S[table-number-alignment=left]
S[table-number-alignment=left]
}
\hline\hline
& \multicolumn{2}{c}{SR-Comp1} & \multicolumn{2}{c}{SR-Comp2} & \multicolumn{2}{c}{SR-Comp3} & \multicolumn{2}{c}{SR-Comp-1c}\\ \hline{$Z$+jets} & \multicolumn{2}{l}{$0.9\phantom{0} \pm 0.4$} & \multicolumn{2}{l}{$1.6\phantom{0} \pm 0.9$} & \multicolumn{2}{l}{$3.1\phantom{0} \pm 0.9$} & \multicolumn{2}{l}{$\phantom{0}4.3\phantom{0} \pm 1.2$}\\
{$W$+jets} & \multicolumn{2}{l}{$2.2\phantom{0} \pm 1.0$} & \multicolumn{2}{l}{$1.3\phantom{0} \pm 0.7$} & \multicolumn{2}{l}{$0.9\phantom{0} \pm 0.6$} & \multicolumn{2}{l}{$\phantom{0}2.8\phantom{0} \pm 0.9$}\\
{Other} & \multicolumn{2}{l}{$0.56 \pm 0.19$} & \multicolumn{2}{l}{$0.62 \pm 0.20$} & \multicolumn{2}{l}{$1.04 \pm 0.34$} & \multicolumn{2}{l}{$\phantom{0}0.68 \pm 0.29$}\\
\hline{Total SM} & \multicolumn{2}{l}{$3.7\phantom{0} \pm 1.2$} & \multicolumn{2}{l}{$3.5\phantom{0} \pm 1.2$} & \multicolumn{2}{l}{$5.0\phantom{0} \pm 1.2$} & \multicolumn{2}{l}{$\phantom{0}7.9\phantom{0} \pm 1.7$}\\ \hline{Observed} & \multicolumn{1}{l}{$3$} &  & \multicolumn{1}{l}{$6$} &  & \multicolumn{1}{l}{$8$} &  & \multicolumn{1}{l}{$12$} & \\
\hline{$\sigPointNoC$=$(600,550)\,\GeV$} & \multicolumn{2}{l}{$6.1\phantom{0} \pm 1.3$} & \multicolumn{1}{l}{} &  & \multicolumn{1}{l}{} &  & \multicolumn{1}{l}{} & \\
\hline{$\sigPointNoC$=$(550,470)\,\GeV$} & \multicolumn{1}{l}{} &  & \multicolumn{2}{l}{$6.0\phantom{0} \pm 0.8$} & \multicolumn{1}{l}{} &  & \multicolumn{1}{l}{} & \\
\hline{$\sigPointNoC$=$(550,375)\,\GeV$} & \multicolumn{1}{l}{} &  & \multicolumn{1}{l}{} &  & \multicolumn{2}{l}{$7.9\phantom{0} \pm 2.1$} & \multicolumn{1}{l}{} & \\
\hline{$\sigPointNoC$=$(450,430)\,\GeV$} & \multicolumn{1}{l}{} &  & \multicolumn{1}{l}{} &  & \multicolumn{1}{l}{} &  & \multicolumn{2}{l}{$11.2\phantom{0} \pm 1.7$}\\
\hline\hline
\end{tabular}

% End of text imported from the .//tex/SRCYieldTable.tex input file

\end{table}
\end{DIFnomarkup}

\begin{DIFnomarkup}
\begin{table}[htbp!]
\caption{Summary of the extracted normalization factors when considering only the likelihood in the control regions for the High-Mass (HM) regions, the Compressed regions that require two $c$-tagged jets (Comp), and the Compressed regions requiring exactly one $c$-tagged jet (Comp-1c). Uncertainties include both the statistical and systematic components but do not include the extrapolation into the signal region, which can reduce the total uncertainty. The $W$+jets normalization factors have larger uncertainties because of the larger effect of theoretical uncertainties. }
\label{tab:norm}
\centering

\begin{tabular}{lc}
\hline\hline
Region  & Normalization factor   \\
\hline
HM $Z$+jets & $1.0\pm0.4$ \\
Comp $Z$+jets & $1.0\pm0.4$ \\
Comp-1c $Z$+jets & $1.1\pm0.4$ \\
HM $W$+jets & $1.4\pm0.6$ \\
Comp $W$+jets & $1.3\pm0.7$ \\
Comp-1c $W$+jets & $1.6\pm0.7$ \\
\hline\hline
\end{tabular}
\end{table}
\end{DIFnomarkup}

\begin{figure}[htb]
\centering
\includegraphics[width=0.68\columnwidth]{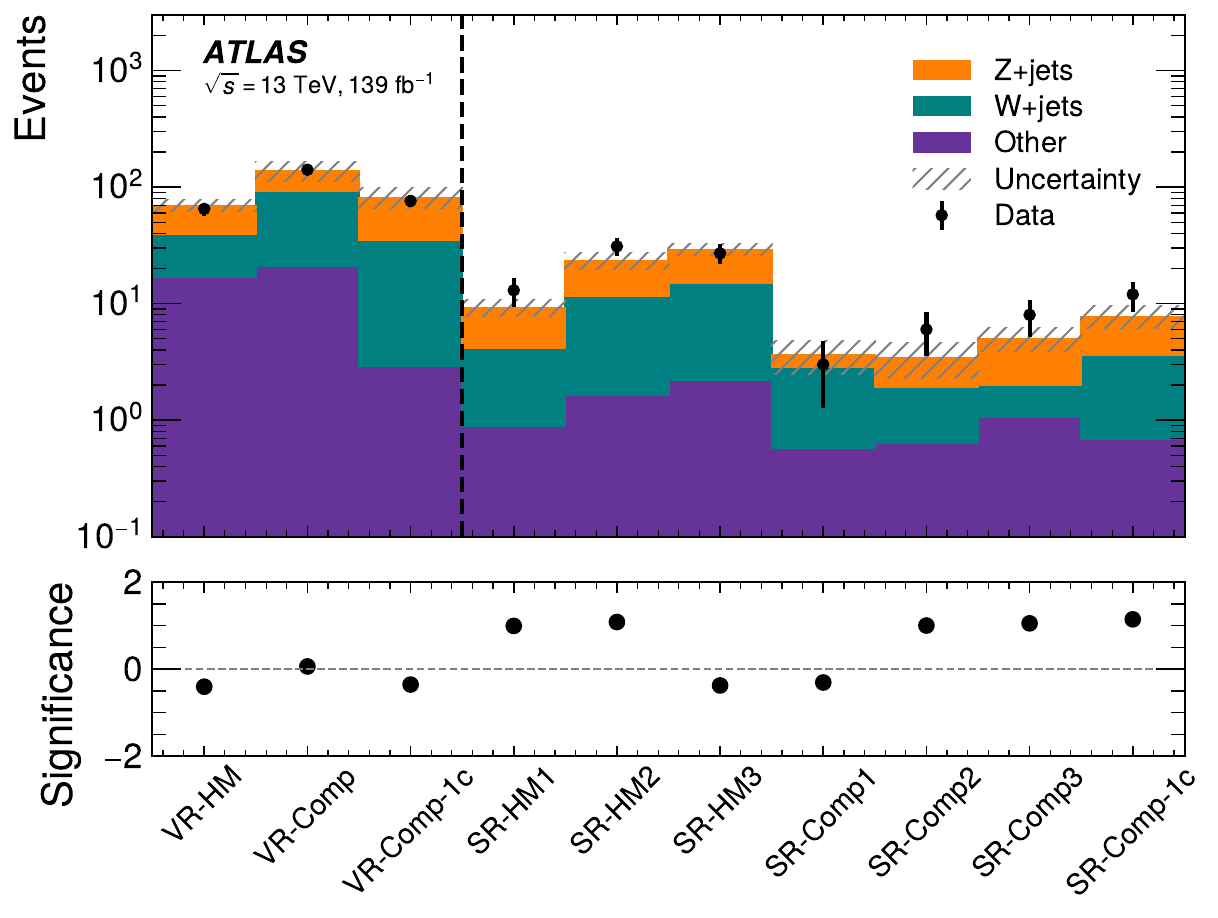}
\caption{Comparison of event yields for the data (points) and SM expectation (stacked histograms) in all
signal and validation regions after the background-only fits. The hatched uncertainty band around the SM prediction
represents the total uncertainty, i.e.\ combining the detector-related and theory-based systematic uncertainties and the MC statistical uncertainties.
The significance of the difference between the data and the SM background prediction, calculated with the profile-likelihood method described in Ref.~\cite{Cousin:2008clt} and using the total uncertainty, is shown in the bottom panel. Processes with top quarks and multiple vector bosons are included in \enquote{Other}.}
\label{fig:srVRSummary}
\end{figure}

\begin{figure}[htb]
\centering
\subfloat[]{\includegraphics[width=0.4\columnwidth]{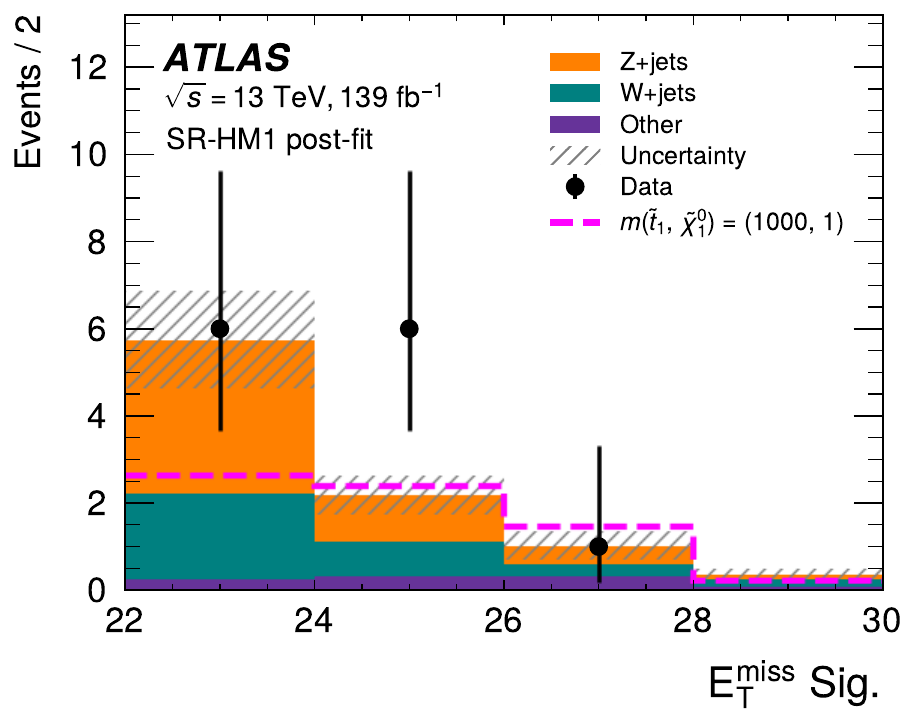}}
\subfloat[]{\includegraphics[width=0.4\columnwidth]{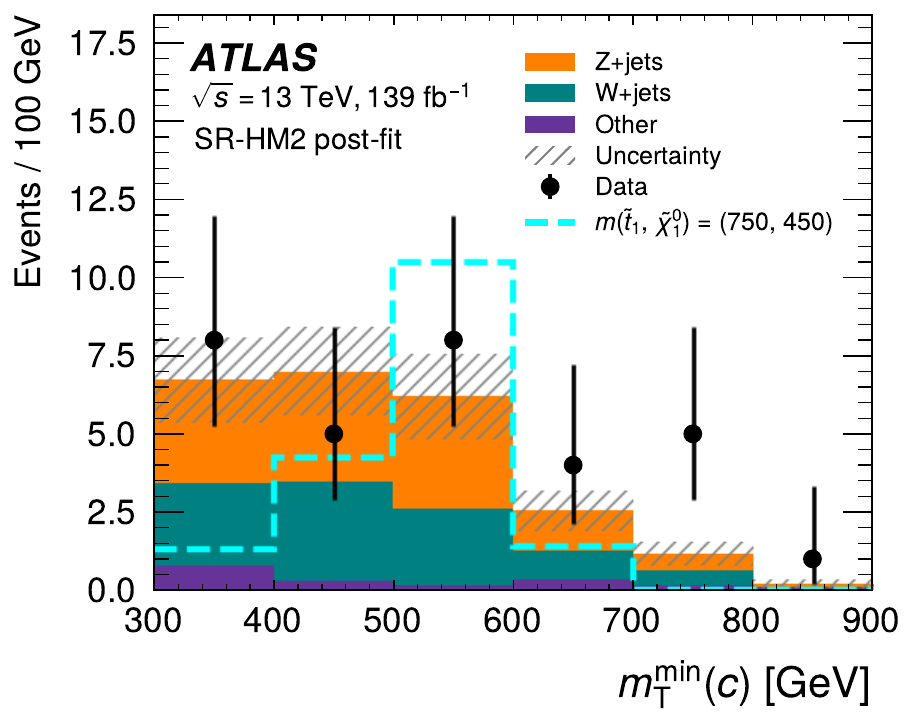}}\\
\subfloat[]{\includegraphics[width=0.4\columnwidth]{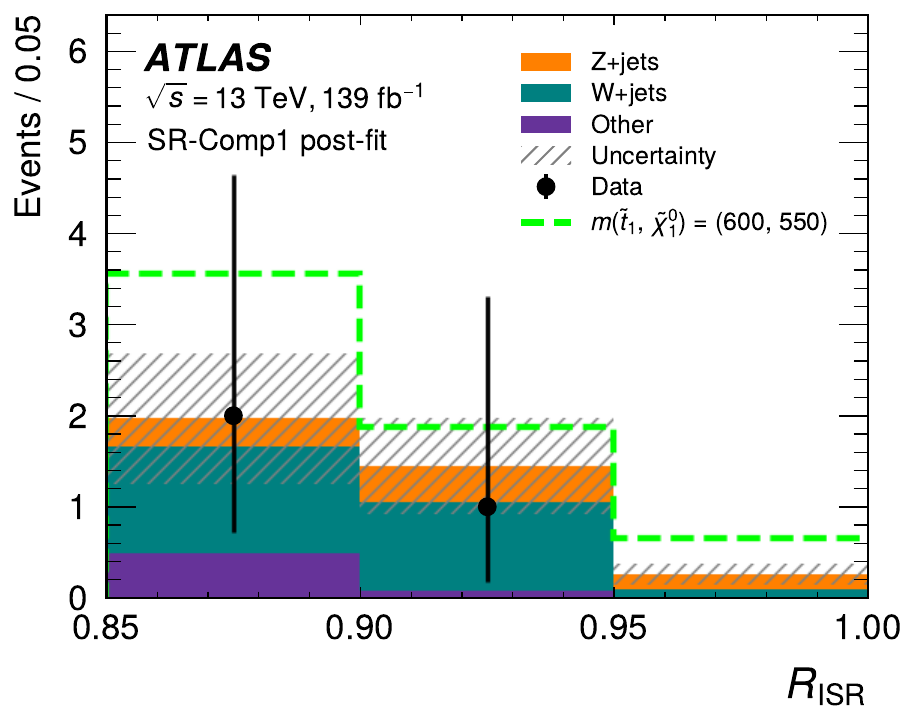}}
\subfloat[]{\includegraphics[width=0.4\columnwidth]{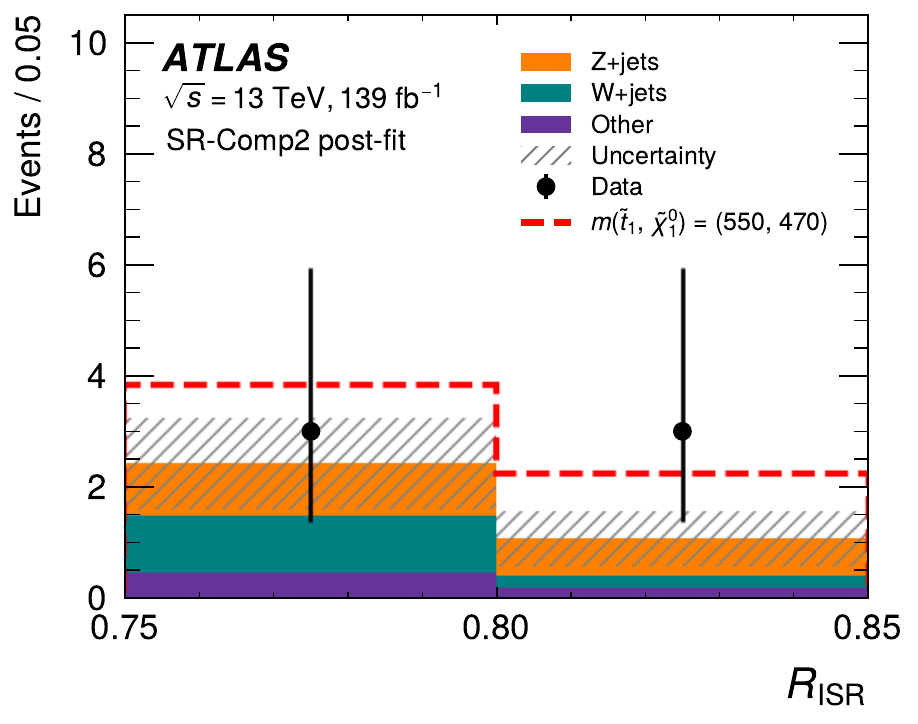}}\\
\subfloat[]{\includegraphics[width=0.4\columnwidth]{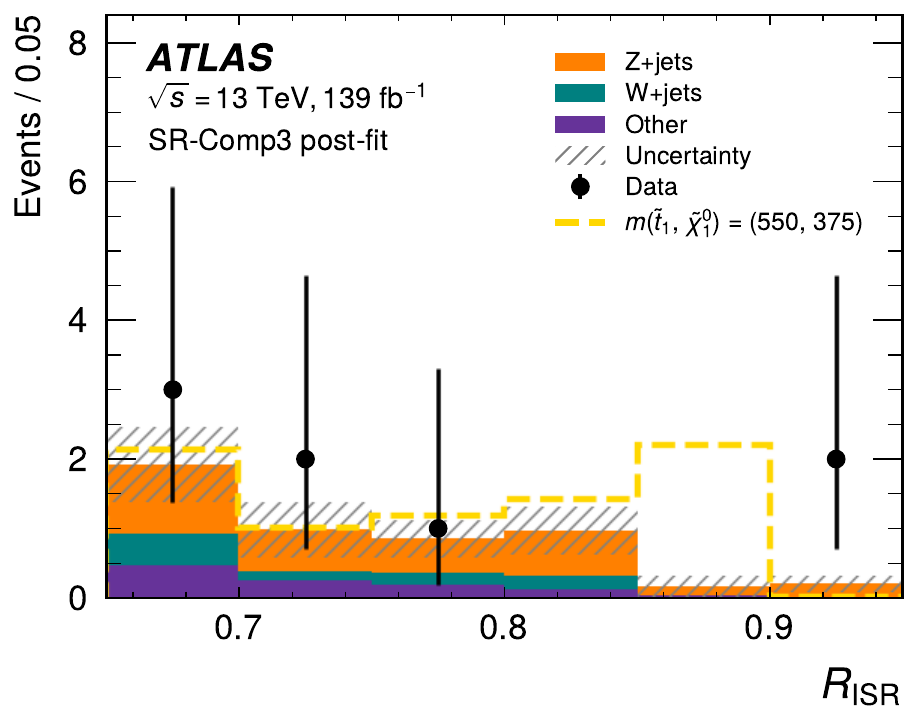}}
\subfloat[]{\includegraphics[width=0.4\columnwidth]{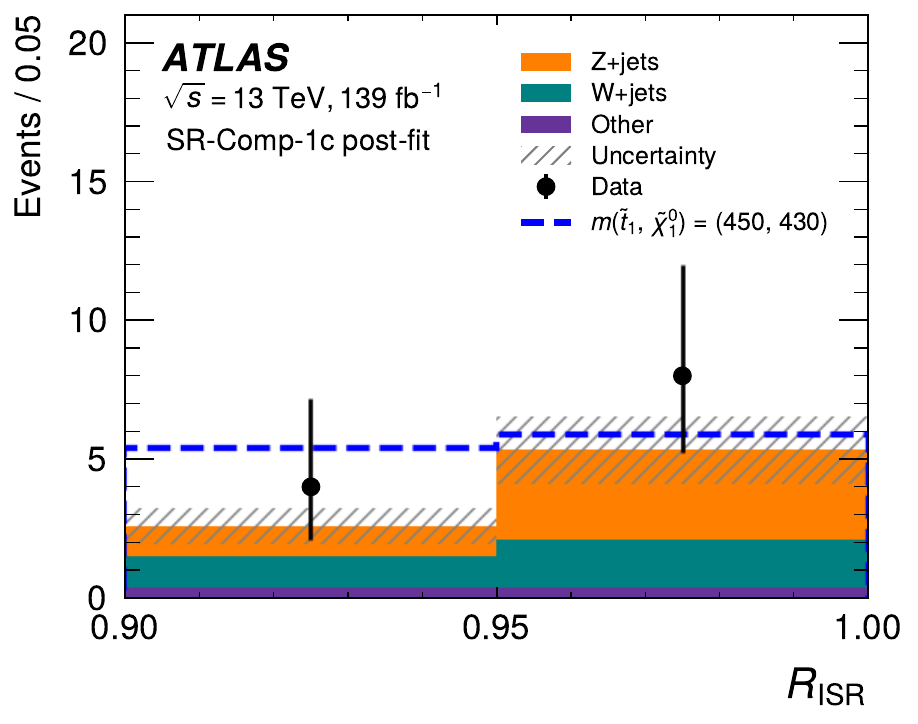}}

\caption{Distributions for the signal regions showing the data (points) and the expected backgrounds (stacked histograms), after simultaneously fitting all the control regions. The
histograms are binned in the key variables for each signal region: (a) \metsig\ for SR-HM1, (b) \mtcmin\ for SR-HM2, (c) \risr\ for SR-Comp1, (d) \risr\ for SR-Comp2, (e) \risr\ for
SR-Comp3, and (f) \risr\ for SR-Comp-1c. The hatched band is the total systematic uncertainty for the expected backgrounds. Distributions of representative $\stopone\overline{\stopone}$
signals for different mass assumptions (or similar) are shown as dashed lines using pre-fit signal contributions. Processes with top quarks and multiple vector bosons are included in
\enquote{Other}. The right-most bin in each histogram does not include the overflow entry but the $x$-axis range is chosen to include all observed data.}

\label{fig:srDists}
\end{figure}

\par The exclusion fit is performed for each signal model individually, using the signal and control regions. For SUSY models, the two statistical combinations considered use the three
High-Mass signal regions with non-overlapping Compressed signal regions: the first combines SR-HM1, SR-HM2, SR-HM3, SR-Comp1, SR-Comp2, and SR-Comp-1c, while the second combines SR-HM1,
SR-HM2, SR-HM3, SR-Comp3, and SR-Comp-1c. For both combinations, all six control regions are used to estimate the background contributions. To maximize the exclusion power of the two
configurations, the combination with the lowest expected CL$_\text{S}$~\cite{Read:2002hq} value is chosen for the final result. The expected sensitivity is calculated by generating
pseudo-experiments that assume an absence of BSM signals. Exclusion limits are set at 95\% confidence level (CL) and are shown for the considered SUSY models in \cref{fig:limits}. The
expected exclusion limits from this search exceed those from the previous ATLAS search that used $c$-tagging~\cite{SUSY-2016-26} (grey-filled contour and dashed line in \cref{fig:limits}),
while being complementary to an ATLAS search for single squark production that did not utilize $c$-tagging~\cite{SUSY-2018-22} (green-filled contour and dashed line in \cref{fig:limits_a})
for $m_{\stopone/\scharmone}\lesssim 920$~\GeV and $\mLSP\gtrsim 400$~\GeV, and is also complementary to an ATLAS single-jet search~\cite{EXOT-2018-06} (cyan-filled contour and dashed line
in \cref{fig:limits_b}) for $\Delta\sigPoint\gtrsim 30$~\GeV.

\begin{figure}[htb]
\centering
\subfloat[]{\label{fig:limits_a}\includegraphics[width=0.48\columnwidth]{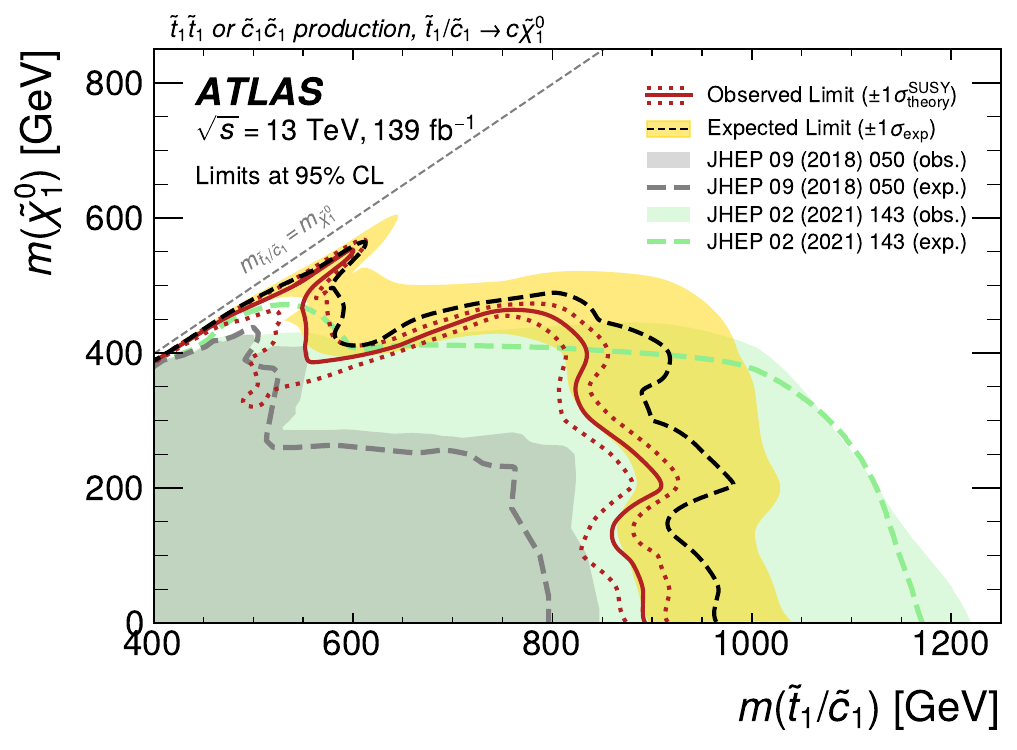}}
\subfloat[]{\label{fig:limits_b}\includegraphics[width=0.48\columnwidth]{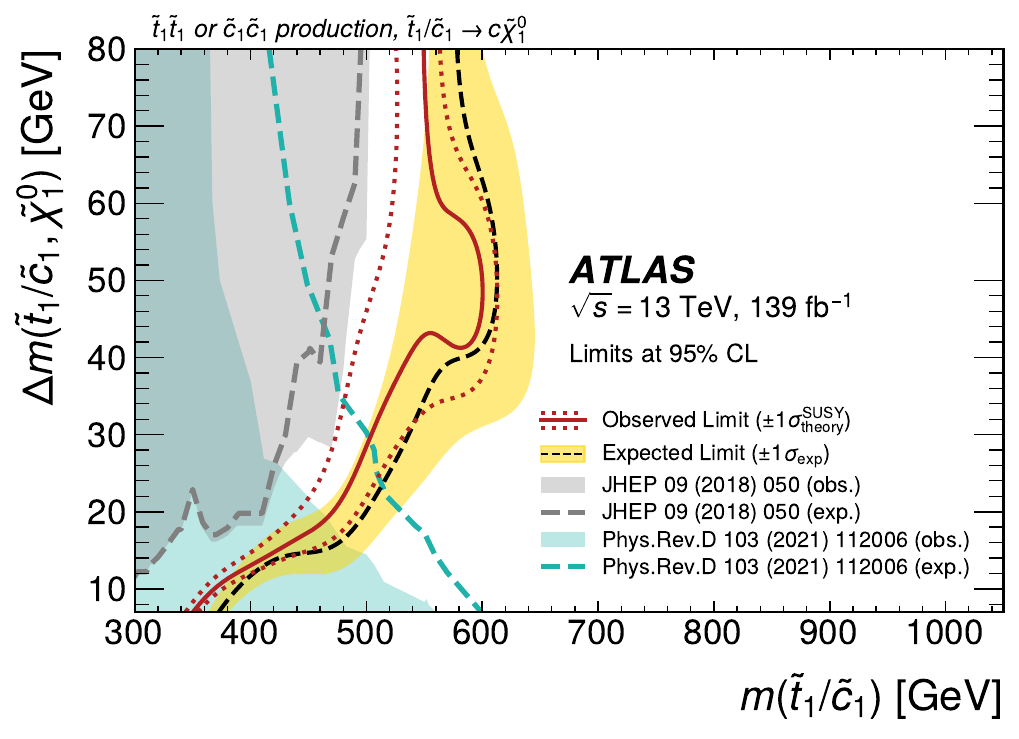}}
\caption{Exclusion limits at 95\% CL for pair production of top
squarks decaying into charm quarks and neutralinos. The expected
limit is shown as a dashed line while the $\pm$1$\sigma$ interval
is shown as a yellow band. The observed limit is shown as a red line and the effect of varying the signal cross-section by $\pm$1$\sigma$ of its predicted theoretical uncertainty is shown as red dashed lines. The contours are shown as two-dimensional projections for masses of squarks and neutralinos (a) and their differences (b). The exclusion contour from the previous ATLAS result that used charm tagging (Ref.~\cite{SUSY-2016-26}) is shown as a grey-dashed line (expected) and grey-filled contour (observed). For (a), the single-squark-production exclusion contour from the ATLAS squarks and gluinos search that did not utilize charm tagging (Ref.~\cite{SUSY-2018-22}) is shown as a green-dashed line (expected) and green-filled contour (observed), while for (b), the contour from the ATLAS single-jet search (Ref.~\cite{EXOT-2018-06}) is shown as a cyan-dashed line (expected) and cyan-filled contour (observed).
}
\label{fig:limits}
\end{figure}

\par Exclusion limits are also presented in \cref{fig:limits-susybr} for an alternative interpretation in which top squarks are pair produced and each decays into a neutralino and either a
top or a charm quark, with $\mathcal{B}(\stopone\to c+\LSP)$ being varied between~0.1 and~1. Due to the $b$-jet veto imposed, none of the signal and control regions are sensitive to the
decay pathway producing a top quark; nonetheless these decay modes are included in the simulated signal samples. Two cases are studied: the mass of the neutralino is fixed to either
\qty{1}{\GeV} or \qty{200}{\GeV}. The strongest limits are set when the top squark decays only via the $c+\LSP$ channel, as expected, with top squark masses up to approximately
\qty{900}{\GeV} being excluded at 95\% CL in both cases. The scenario in which $\mLSP=\qty{1}{\GeV}$ presents a boosted decay topology over the entire range of top squark masses considered,
and as such, the major constraining power comes from the High-Mass signal regions. A monotonic decrease in sensitivity is observed with decreasing branching fraction to $c+\LSP$. For
$\mLSP=\qty{200}{\GeV}$ neutralinos, the Compressed signal regions gradually become more sensitive as the top squark mass decreases, causing the change in behaviour near
$m(\stopone)=\qty{500}{\GeV}$.

\begin{figure}[htb]
\centering
\subfloat[]{\includegraphics[width=0.48\columnwidth]{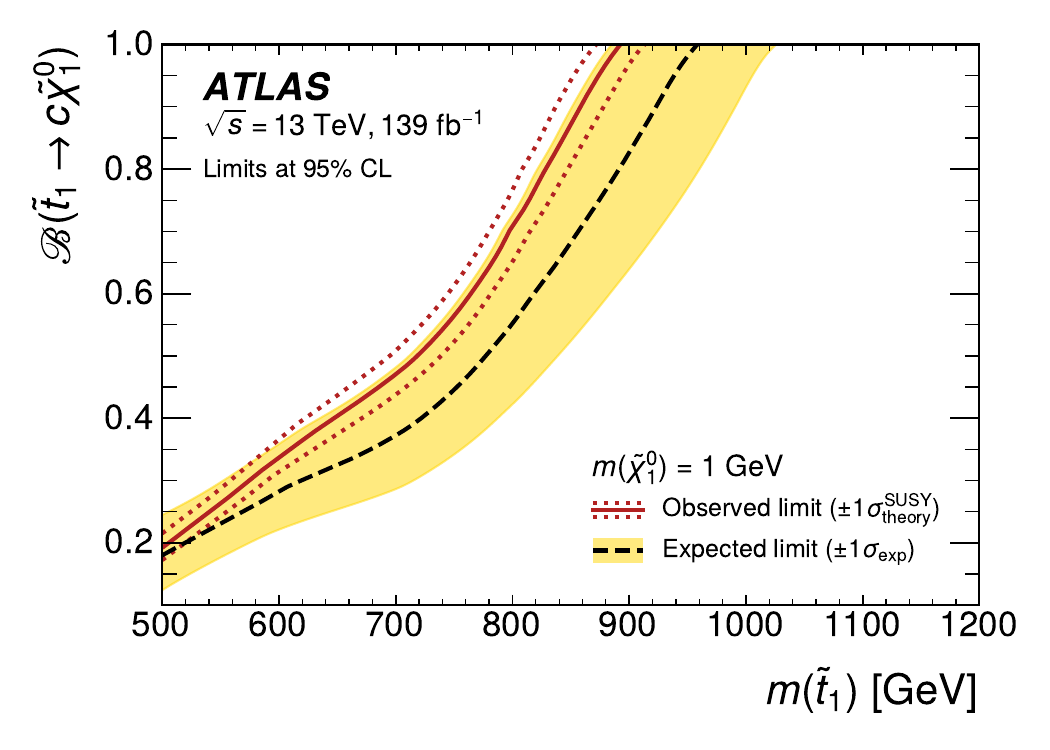}}
\subfloat[]{\includegraphics[width=0.48\columnwidth]{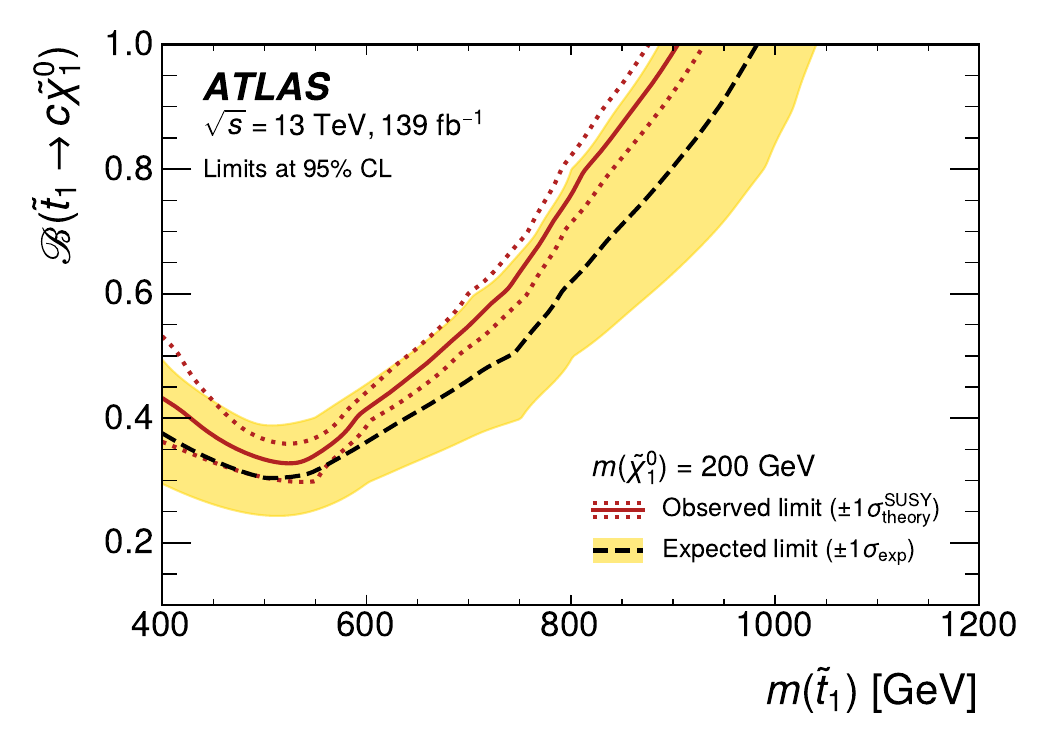}}
\caption{The expected and observed exclusion contours for pair
production of top squarks decaying into top or charm quarks and
neutralinos. The contours are shown as two-dimensional projections
in the plane of the squark mass and assumed branching fraction,
whose value is varied between~0~(all decays to $t+\LSP$)
and~1~(all decays to $c+\LSP$). The mass of the neutralino is
fixed to either (a) \qty{1}{\GeV} or (b) \qty{200}{\GeV}. The
cross-section limits for the contours are calculated by combining
the signal regions as discussed earlier in the paper. The
branching fractions above and masses below the lines (high
branching fraction and low mass, and thus high production
cross-section) are excluded. The expected limit is shown as a
dashed line, while the $\pm$1$\sigma$ interval is shown as a yellow band. The observed limit is shown as a red line and the effect of varying the signal cross-section by $\pm$1$\sigma$ of its predicted theoretical uncertainty is shown as red dashed lines.}
\label{fig:limits-susybr}
\end{figure}

\par Additional limits are set in terms of LQ models after performing a combined fit of the LQ signals using only the High-Mass regions because of similarities in kinematics in the scenario
where the LSP is massless. In the case of scalar-LQ models, events were reweighted to span the full range of branching fractions of the LQ decay into $c\nu$. Expected and observed exclusion
limits on the $\mathcal{B}(\lqu\to c\nu_{e})$ and $\mathcal{B}(\lqu\to c\nu_{\mu})$ at 95\% CL are presented in \cref{fig:HighMass_BestExpected_Exclusion_sLQ} as a function of the
leptoquark mass, with the observed upper limit on the production cross-section shown on the $z$-axis. Leptoquarks with masses up to approximately \qty{900}{\GeV} are excluded.

\begin{figure}[htb]
\centering
\subfloat[]{\includegraphics[width=0.5\columnwidth]{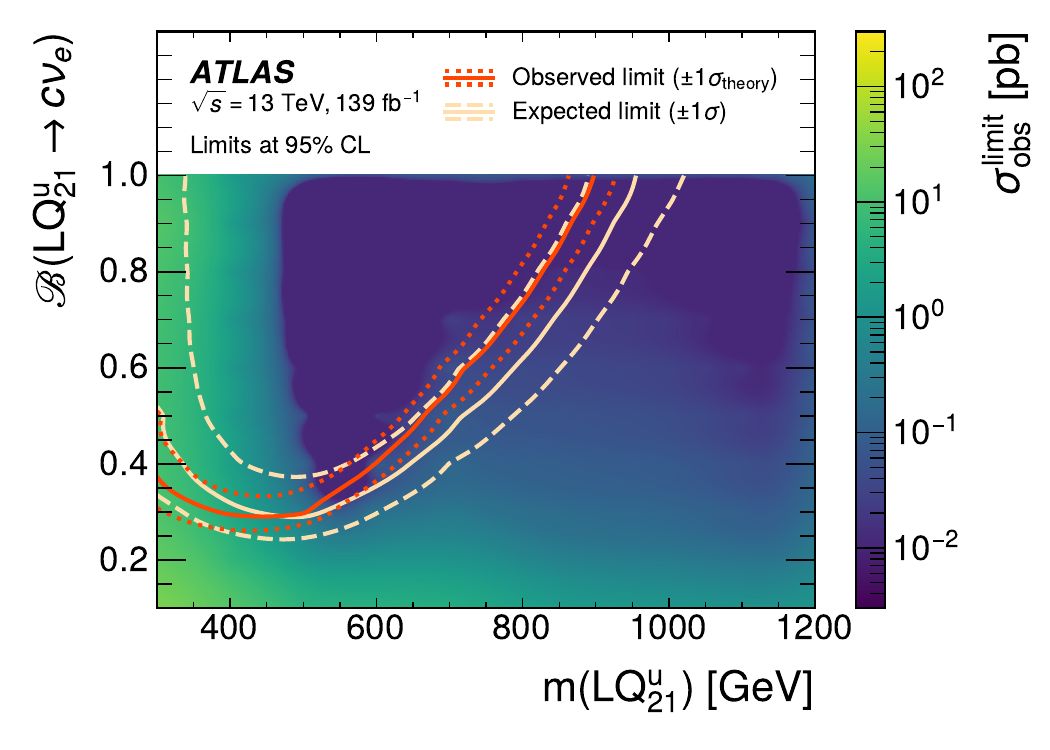}}%
\subfloat[]{\includegraphics[width=0.5\columnwidth]{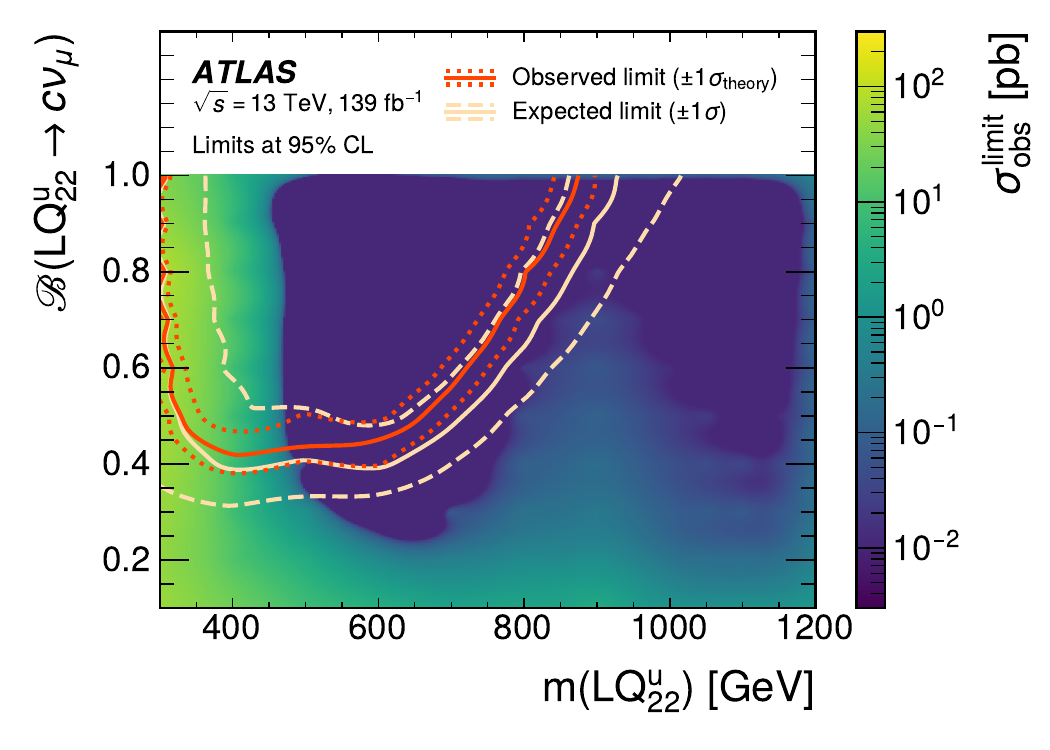}}

\caption{Expected (white) and observed (red) exclusion limits at
95\% CL for up-type scalar LQs coupled to (a) first- and (b)
second-generation leptons. The observed upper limit on the
production cross-section is plotted on the $z$-axis as a function
of (a) $\mathcal{B}(\lqu\to c\nu_{e})$ with the alternative decay
being $\lqu\to\ s+e$ and (b) $\mathcal{B}(\lqu\to c\nu_{\mu})$ with the alternative decay
being $\lqu\to\ s+\mu$  ($y$-axis) and the leptoquark masses ($x$-axis). The branching fractions above and masses below the lines (high branching fraction and low mass, and
thus high production cross-section) are excluded. The $\pm$1$\sigma$ interval of the expected limit is shown as dashed white lines while the effect of varying the signal cross-section by $\pm$1$\sigma$ of its predicted theoretical uncertainty is shown as red dashed lines. }

\label{fig:HighMass_BestExpected_Exclusion_sLQ}
\end{figure}

\par Vector-LQ models are excluded at 95\% CL up to masses of \qty{950}{\GeV} in the minimal coupling scenario,
and \qty{1150}{\GeV} for the Yang--Mills case,
for an assumed branching fraction of $\mathcal{B}(\vlq\to c\nu_\tau)=0.5$.
Upper limits on the production cross-section for both the minimal and
Yang--Mills coupling scenarios are shown in \cref{fig:HighMass_BestExpected_UpperLimits_vLQ},
with the theoretical cross-section at LO and the $\pm1\sigma$ interval
(encompassing the effects of PDF, \alphas, renormalization and factorization scale variations) shown in blue.

\begin{figure}[htb]
\centering
\subfloat[]{\includegraphics[width=0.5\textwidth]{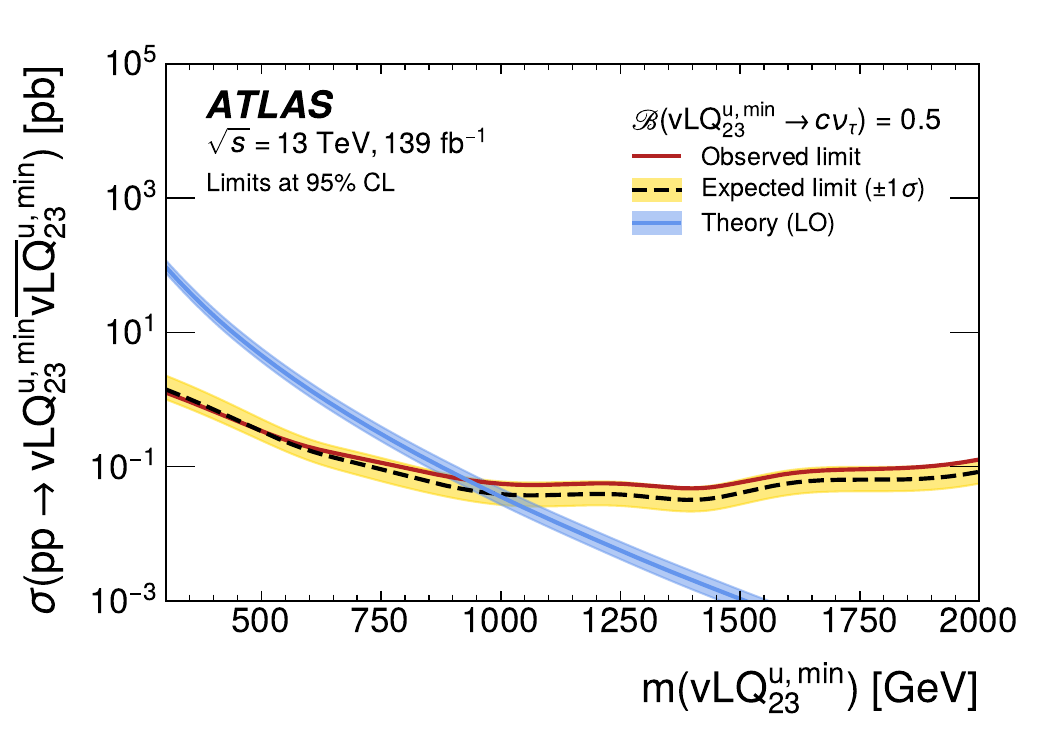}}%
\subfloat[]{\includegraphics[width=0.5\textwidth]{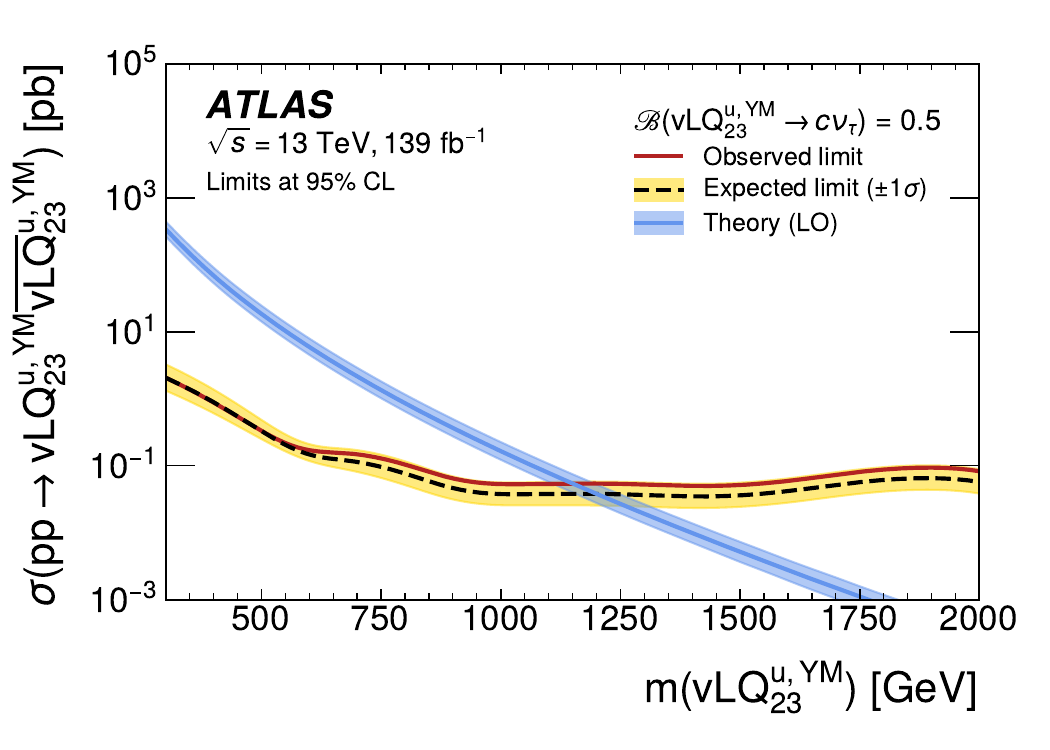}}
\caption{Expected (dashed) and observed (solid) cross-section upper limits as a function of leptoquark mass for $U(1)$ vector-LQ models in the (a) minimal and (b) Yang--Mills coupling scenarios
with $\beta_{23}=1$, corresponding to a branching fraction $\mathcal{B}(\vlq\to c\nu_\tau)=0.5$. The $\pm$1$\sigma$ interval of the expected limit is shown as a yellow band.
The theoretical prediction and its $\pm1\sigma$ interval band are shown in blue.}
\label{fig:HighMass_BestExpected_UpperLimits_vLQ}
\end{figure}

\par A model-independent fit is performed for signal regions with the least stringent requirements: SR-HM1, SR-Comp3, and SR-HM-Disc. For the model-independent interpretation, the SM background estimates are fit to the observed yields in both the CRs and SRs, testing for the potential presence of any BSM events in the SRs. The observed cross-section limit, $\langle \epsilon\sigma \rangle_{\mathrm{obs}}^{95}$ , is calculating by dividing the
observed limit on the signal strength, $S_{\mathrm{obs}}^{95}$, by the integrated luminosity. These limits on the number of BSM events are shown in \cref{tab:discovery_fit_results} and can be reinterpreted for a generic BSM model by calculating the identification
efficiency and detector acceptance for the model of interest.

\begin{table}[htb]
\begin{center}
\def\arraystretch{1.5}
\caption{Model-independent upper limits on the BSM event yields and cross-sections. The $\langle \epsilon\sigma \rangle_{\mathrm{obs}}^{95}$ is the observed upper limit on the visible BSM
cross-section at 95\% CL, and $S_{\mathrm{obs}}^{95}$ is the observed upper limit on the number of BSM events. Similarly, $S_{\mathrm{exp}}^{95}$ is the expected upper limit on the number
of BSM events but it is computed  assuming that the observed number of events is identical to the number of background events. The discovery $p$-value ($p_{\mathrm{obs}}(S=0)$), which
tests the compatibility of the observed data with the background-only (zero signal yield) hypothesis relative to fluctuations of the background, is shown in the last column.}

\label{tab:discovery_fit_results}
\begin{tabular}{lcccc}
\hline\hline
Signal Region          & $\langle \epsilon\sigma \rangle_{\mathrm{obs}}^{95}$ [fb] & $S_{\mathrm{obs}}^{95}$   & $S_{\mathrm{exp}}^{95}$ & $p_{\mathrm{obs}}(S=0)$      \\
\hline
SR-HM1                    & 0.09                                                      & 12.8                     & 9.7$^{+4.2}_{-2.9}$      &  0.19 \\
SR-HM-Disc               & 0.36                                                      & 50.7                     & 34$^{+13}_{-9}$   &  0.09 \\
SR-Comp3                   & 0.07                                                      & ~~9.2                      & 6.5$^{+3.2}_{-2.0}$      &  0.15 \\
SR-Comp-1c                    & 0.08                                                      & 11.6                     & 8.0$^{+3.8}_{-2.4}$      &  0.13 \\
\hline\hline
\end{tabular}
\end{center}

\end{table}

\FloatBarrier
\clearpage
\section{Conclusion}

\par This paper presents a search targeting BSM models contributing to final states with missing transverse momentum and charm-tagged jets, based on a $\qty{139}{\ifb}$ dataset of
proton--proton collisions at $\sqs=\qty{13}{\TeV}$ recorded by the ATLAS experiment at the LHC. The observed experimental data are found to agree with the SM background estimate. The results
are interpreted as upper limits at 95\% CL on the production cross-section of selected SUSY and leptoquark models and as model-independent upper limits on the production cross-section for
BSM particles. Top/charm squark masses are excluded up to ${\sim}900$~\GeV, assuming each squark decays into a charm quark and an LSP. This is a ${\sim}100$~\GeV improvement on the previous
ATLAS search that used $c$-tagging. The increase in sensitivity is driven by the larger experimental dataset, the sophisticated design of the signal regions (which includes different
strategies for large and small mass splittings between the top squark and the LSP), and improvements in $c$-tagging. The search also excludes, for the first time at the LHC, scalar
leptoquarks with a mass below 900~\GeV that decay into a charm quark and either a first- or second-generation neutrino. Finally, vector-leptoquark models in the minimal and Yang--Mills
coupling scenarios are excluded at 95\% CL up to masses of 950~\GeV and 1150~\GeV, respectively, assuming a branching fraction of $\mathcal{B}(\vlq\to c\nu_\tau)=0.5$.

% End of text imported from the .//tex/results_conclusion.tex input file

%
\section*{Acknowledgements}
%

% The next lines are included from the .//acknowledgements/Acknowledgements.tex input file
%
%

%
%

We thank CERN for the very successful operation of the LHC and its injectors, as well as the support staff at
CERN and at our institutions worldwide without whom ATLAS could not be operated efficiently.

The crucial computing support from all WLCG partners is acknowledged gratefully, in particular from CERN, the ATLAS Tier-1 facilities at TRIUMF/SFU (Canada), NDGF (Denmark, Norway, Sweden), CC-IN2P3 (France), KIT/GridKA (Germany), INFN-CNAF (Italy), NL-T1 (Netherlands), PIC (Spain), RAL (UK) and BNL (USA), the Tier-2 facilities worldwide and large non-WLCG resource providers. Major contributors of computing resources are listed in Ref.~\cite{ATL-SOFT-PUB-2023-001}.

We gratefully acknowledge the support of ANPCyT, Argentina; YerPhI, Armenia; ARC, Australia; BMWFW and FWF, Austria; ANAS, Azerbaijan; CNPq and FAPESP, Brazil; NSERC, NRC and CFI, Canada; CERN; ANID, Chile; CAS, MOST and NSFC, China; Minciencias, Colombia; MEYS CR, Czech Republic; DNRF and DNSRC, Denmark; IN2P3-CNRS and CEA-DRF/IRFU, France; SRNSFG, Georgia; BMBF, HGF and MPG, Germany; GSRI, Greece; RGC and Hong Kong SAR, China; ISF and Benoziyo Center, Israel; INFN, Italy; MEXT and JSPS, Japan; CNRST, Morocco; NWO, Netherlands; RCN, Norway; MNiSW, Poland; FCT, Portugal; MNE/IFA, Romania; MESTD, Serbia; MSSR, Slovakia; ARIS and MVZI, Slovenia; DSI/NRF, South Africa; MICIU/AEI, Spain; SRC and Wallenberg Foundation, Sweden; SERI, SNSF and Cantons of Bern and Geneva, Switzerland; NSTC, Taipei; TENMAK, T\"urkiye; STFC/UKRI, United Kingdom; DOE and NSF, United States of America.

Individual groups and members have received support from BCKDF, CANARIE, CRC and DRAC, Canada; CERN-CZ, FORTE and PRIMUS, Czech Republic; COST, ERC, ERDF, Horizon 2020, ICSC-NextGenerationEU and Marie Sk{\l}odowska-Curie Actions, European Union; Investissements d'Avenir Labex, Investissements d'Avenir Idex and ANR, France; DFG and AvH Foundation, Germany; Herakleitos, Thales and Aristeia programmes co-financed by EU-ESF and the Greek NSRF, Greece; BSF-NSF and MINERVA, Israel; Norwegian Financial Mechanism 2014-2021, Norway; NCN and NAWA, Poland; La Caixa Banking Foundation, CERCA Programme Generalitat de Catalunya and PROMETEO and GenT Programmes Generalitat Valenciana, Spain; G\"{o}ran Gustafssons Stiftelse, Sweden; The Royal Society and Leverhulme Trust, United Kingdom.

In addition, individual members wish to acknowledge support from CERN: European Organization for Nuclear Research (CERN PJAS); Chile: Agencia Nacional de Investigaci\'on y Desarrollo (FONDECYT 1190886, FONDECYT 1230812, FONDECYT 1230987); China: Chinese Ministry of Science and Technology (MOST-2023YFA1605700, MOST-2023YFA1609300), National Natural Science Foundation of China (NSFC - 12175119, NSFC 12275265, NSFC-12075060); Czech Republic: Czech Science Foundation (GACR - 24-11373S), Ministry of Education Youth and Sports (FORTE CZ.02.01.01/00/22\_008/0004632), PRIMUS Research Programme (PRIMUS/21/SCI/017); EU: H2020 European Research Council (ERC - 101002463); European Union: European Research Council (ERC - 948254, ERC 101089007), Horizon 2020 Framework Programme (MUCCA - CHIST-ERA-19-XAI-00), European Union, Future Artificial Intelligence Research (FAIR-NextGenerationEU PE00000013), Italian Center for High Performance Computing, Big Data and Quantum Computing (ICSC, NextGenerationEU); France: Agence Nationale de la Recherche (ANR-20-CE31-0013, ANR-21-CE31-0013, ANR-21-CE31-0022, ANR-22-EDIR-0002), Investissements d'Avenir Labex (ANR-11-LABX-0012); Germany: Baden-Württemberg Stiftung (BW Stiftung-Postdoc Eliteprogramme), Deutsche Forschungsgemeinschaft (DFG - 469666862, DFG - CR 312/5-2); Italy: Istituto Nazionale di Fisica Nucleare (ICSC, NextGenerationEU), Ministero dell'Università e della Ricerca (PRIN - 20223N7F8K - PNRR M4.C2.1.1); Japan: Japan Society for the Promotion of Science (JSPS KAKENHI JP21H05085, JSPS KAKENHI JP22H01227, JSPS KAKENHI JP22H04944, JSPS KAKENHI JP22KK0227); Netherlands: Netherlands Organisation for Scientific Research (NWO Veni 2020 - VI.Veni.202.179); Norway: Research Council of Norway (RCN-314472); Poland: Ministry of Science and Higher Education (IDUB AGH, POB8, D4 no 9722), Polish National Agency for Academic Exchange (PPN/PPO/2020/1/00002/U/00001), Polish National Science Centre (NCN 2021/42/E/ST2/00350, NCN OPUS nr 2022/47/B/ST2/03059, NCN UMO-2019/34/E/ST2/00393, NCN \& H2020 MSCA 945339, UMO-2020/37/B/ST2/01043, UMO-2021/40/C/ST2/00187, UMO-2022/47/O/ST2/00148, UMO-2023/49/B/ST2/04085); Slovenia: Slovenian Research Agency (ARIS grant J1-3010); Spain: Generalitat Valenciana (Artemisa, FEDER, IDIFEDER/2018/048), Ministry of Science and Innovation (MCIN \& NextGenEU PCI2022-135018-2, MICIN \& FEDER PID2021-125273NB, RYC2019-028510-I, RYC2020-030254-I, RYC2021-031273-I, RYC2022-038164-I), PROMETEO and GenT Programmes Generalitat Valenciana (CIDEGENT/2019/023, CIDEGENT/2019/027); Sweden: Carl Trygger Foundation (Carl Trygger Foundation CTS 22:2312), Swedish Research Council (Swedish Research Council 2023-04654, VR 2018-00482, VR 2022-03845, VR 2022-04683, VR 2023-03403, VR grant 2021-03651), Knut and Alice Wallenberg Foundation (KAW 2018.0157, KAW 2018.0458, KAW 2019.0447, KAW 2022.0358); Switzerland: Swiss National Science Foundation (SNSF - PCEFP2\_194658); United Kingdom: Leverhulme Trust (Leverhulme Trust RPG-2020-004), Royal Society (NIF-R1-231091); United States of America: U.S. Department of Energy (ECA DE-AC02-76SF00515), Neubauer Family Foundation.

%
%

% End of text imported from the .//acknowledgements/Acknowledgements.tex input file

%
\appendix
\printbibliography
\clearpage
% ATLAS Collaboration author list
% Reference date of SUSY-2018-25 is 2024-03-12
% Author list last updated on date 22-JAN-25
% Data extracted on 22-Jan-2025 for paper reference SUSY-2018-25
% at 1:28am
 
\begin{flushleft}
\hypersetup{urlcolor=black}
{\Large The ATLAS Collaboration}

\bigskip

\AtlasOrcid[0000-0002-6665-4934]{G.~Aad}$^\textrm{\scriptsize 104}$,
\AtlasOrcid[0000-0001-7616-1554]{E.~Aakvaag}$^\textrm{\scriptsize 17}$,
\AtlasOrcid[0000-0002-5888-2734]{B.~Abbott}$^\textrm{\scriptsize 123}$,
\AtlasOrcid[0000-0002-0287-5869]{S.~Abdelhameed}$^\textrm{\scriptsize 119a}$,
\AtlasOrcid[0000-0002-1002-1652]{K.~Abeling}$^\textrm{\scriptsize 56}$,
\AtlasOrcid[0000-0001-5763-2760]{N.J.~Abicht}$^\textrm{\scriptsize 50}$,
\AtlasOrcid[0000-0002-8496-9294]{S.H.~Abidi}$^\textrm{\scriptsize 30}$,
\AtlasOrcid[0009-0003-6578-220X]{M.~Aboelela}$^\textrm{\scriptsize 45}$,
\AtlasOrcid[0000-0002-9987-2292]{A.~Aboulhorma}$^\textrm{\scriptsize 36e}$,
\AtlasOrcid[0000-0001-5329-6640]{H.~Abramowicz}$^\textrm{\scriptsize 155}$
% ...
\bigskip
\\

% ...
$^{17}$Department for Physics and Technology, University of Bergen, Bergen; Norway.\\
% ...
$^{30}$Physics Department, Brookhaven National Laboratory, Upton NY; United States of America.\\
% ...
$^{36}$$^{(a)}$Facult\'e des Sciences Ain Chock, Universit\'e Hassan II de Casablanca;$^{(b)}$Facult\'{e} des Sciences, Universit\'{e} Ibn-Tofail, K\'{e}nitra;$^{(c)}$Facult\'e des Sciences Semlalia, Universit\'e Cadi Ayyad, LPHEA-Marrakech;$^{(d)}$LPMR, Facult\'e des Sciences, Universit\'e Mohamed Premier, Oujda;$^{(e)}$Facult\'e des sciences, Universit\'e Mohammed V, Rabat;$^{(f)}$Institute of Applied Physics, Mohammed VI Polytechnic University, Ben Guerir; Morocco.\\
% ...
$^{45}$Physics Department, Southern Methodist University, Dallas TX; United States of America.\\
% ...
$^{50}$Fakult\"{a}t Physik , Technische Universit{\"a}t Dortmund, Dortmund; Germany.\\
% ...
$^{56}$II. Physikalisches Institut, Georg-August-Universit\"{a}t G\"ottingen, G\"ottingen; Germany.\\
% ...
$^{104}$CPPM, Aix-Marseille Universit\'e, CNRS/IN2P3, Marseille; France.\\
% ...
$^{119}$$^{(a)}$New York University Abu Dhabi, Abu Dhabi;$^{(b)}$United Arab Emirates University, Al Ain; United Arab Emirates.\\
% ...
$^{123}$Homer L. Dodge Department of Physics and Astronomy, University of Oklahoma, Norman OK; United States of America.\\
% ...
$^{155}$Raymond and Beverly Sackler School of Physics and Astronomy, Tel Aviv University, Tel Aviv; Israel.\\
% ...
$^{a}$ Also Affiliated with an institute covered by a cooperation agreement with CERN.\\
$^{b}$ Also at An-Najah National University, Nablus; Palestine.\\
$^{c}$ Also at Borough of Manhattan Community College, City University of New York, New York NY; United States of America.\\
$^{d}$ Also at Center for High Energy Physics, Peking University; China.\\
$^{e}$ Also at Center for Interdisciplinary Research and Innovation (CIRI-AUTH), Thessaloniki; Greece.\\
$^{f}$ Also at Centro Studi e Ricerche Enrico Fermi; Italy.\\
% ...

\end{flushleft}

% Created with Glance <Atlas.Glance@cern.ch>

%

%
%

%

%
%
%
%

\end{document}